\documentclass[12pt,letterpaper,hyper]{JHEP3}
\pdfoutput=1

\usepackage[utf8]{inputenc}
\usepackage{epic,eepic}
\usepackage{amsmath,epsfig}
\usepackage{amssymb,amsfonts}
\newcommand{\beq}{\begin{equation}}
\newcommand{\eeq}{\end{equation}}

\def\d{{\rm d}}

\title{Polchinski-Strassler does not uplift Klebanov-Strassler}

\author{Iosif Bena, Mariana Gra\~na, Stanislav Kuperstein and Stefano Massai \\

\it{Institut de Physique Th\'eorique, CEA
Saclay, CNRS URA 2306 \\ F-91191 Gif-sur-Yvette, France} \\

\texttt{e-mails}: \textsf{iosif.bena, mariana.grana, stanislav.kuperstein, stefano.massai@cea.fr}

}

\abstract{ Anti-D3-branes at the tip of the Klebanov-Strassler solution with D3-charge dissolved in fluxes give rise, in the probe approximation, to a metastable state. The fully back-reacted smeared solution has singular three-form fluxes in the IR, whose presence suggests a stringy resolution by brane polarization \`a la Polchinski-Strassler. In this paper we show that there is no polarization into anti-D5-branes wrapping the $S^2$ of the conifold at a finite radius. The singularities therefore do not seem to be physical, signaling that antibranes cannot be used to uplift AdS and obtain a very large landscape of de Sitter vacua in string theory.}

\preprint{IPhT-T12/137}

\begin{document}

\section{Introduction}

An extensive body of work has dealt over the past few years with the important question of the backreaction of anti D-branes in backgrounds that have D-brane charge dissolved in supergravity fluxes, focusing in particular on anti-D3 branes in the Klebanov-Strassler (KS) warped deformed conifold solution \cite{Klebanov:2000hb}. These anti-D3 branes appear to give rise to metastable vacua in the probe approximation \cite{Kachru:2002gs}, but upon taking into account their backreaction at first order \cite{Bena:2009xk,Bena:2011hz, Bena:2011wh}, one finds that their supergravity solution has a certain singularity in the infrared corresponding to a divergence of the energy density in the RR and NSNS three-forms. This singularity has been argued to go away when considering the full backreaction of the anti-D3 branes \cite{Dymarsky:2011pm}, but this hope was short-lived: the fully-backreacted solution describing the infrared of the anti-D3 branes is also singular  \cite{Bena:2012bk}.

Singularities in string theory have been studied extensively over more than ten years, and there are two very important lessons that have come out of this study: the first is that if a solution has a singularity one cannot hope to obtain correct physics by doing calculations in some region far away from the singularity, where the curvature is low; the resolution of the singularity may involve low-mass modes that modify the spacetime at macroscopic distances away from the singularity, or may signal an instability of the whole spacetime.  A second lesson, which is a corollary of the first, is that in the context of the $AdS$-CFT correspondence only singularity-free solutions are dual to vacua of the gauge theory, while singular solutions (such as the Polchinski-Strassler unpolarized solution \cite{Girardello:1999bd,Polchinski:2000uf}, the singular giant graviton \cite{McGreevy:2000cw,Lin:2004nb} or the Klebanov-Tseytlin solution \cite{Klebanov:2000nc}) are not dual to any vacuum of the gauge theory and have to be discarded as unphysical.\footnote{Another possibility is that the singularity is resolved in a manner that we do not understand, but then one expects that it could be cloaked by a horizon if one increases the temperature. In \cite{Bena:2012ek} it has been explicitly shown that this is not the case either.}

\begin{figure}[t]
\begin{center}$
\begin{array}{cc}
\includegraphics[trim = 0mm 150mm 0mm 15mm, clip, width=7cm]{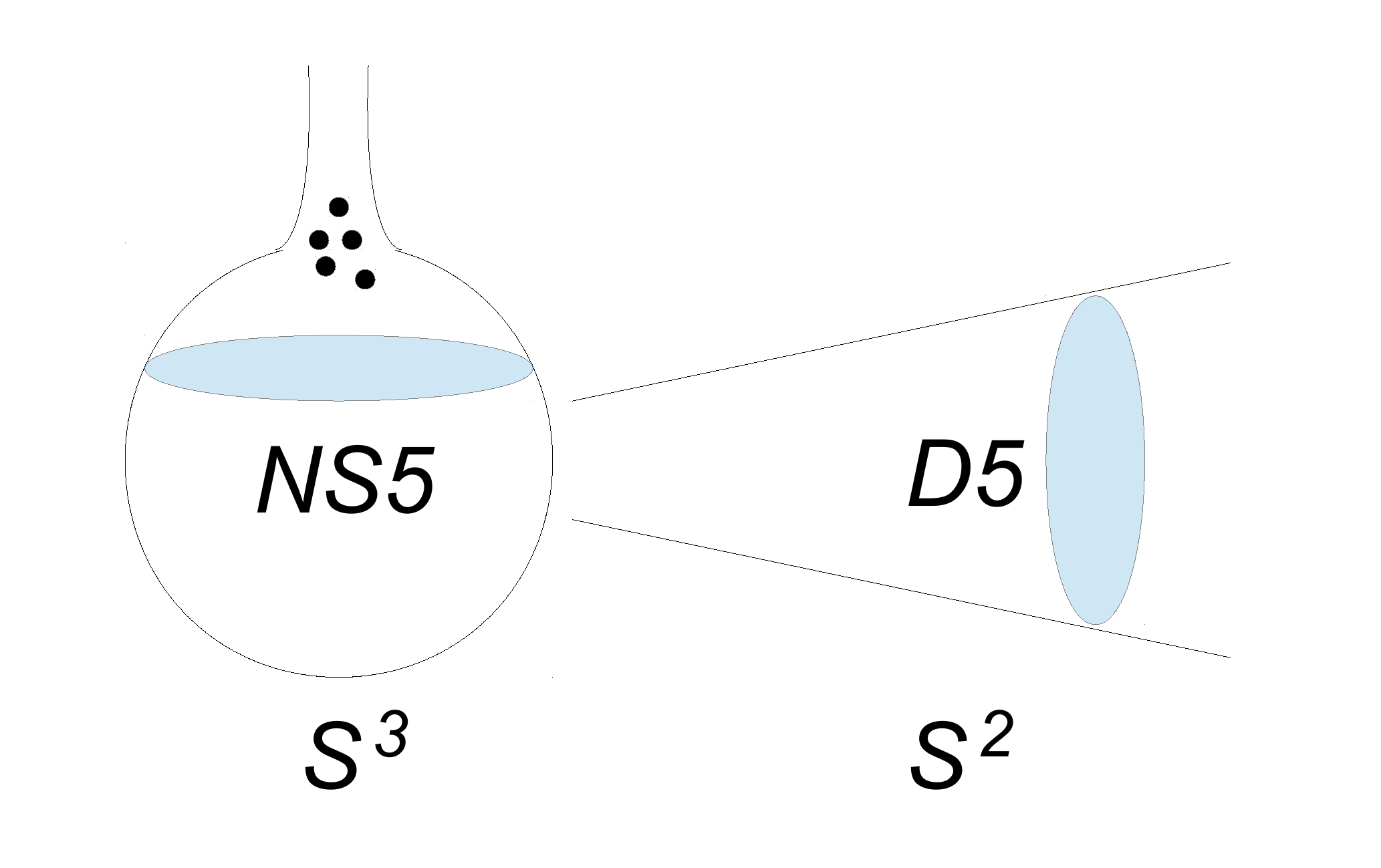} &
\includegraphics[trim = 0mm 150mm 0mm 15mm, clip, width=7cm]{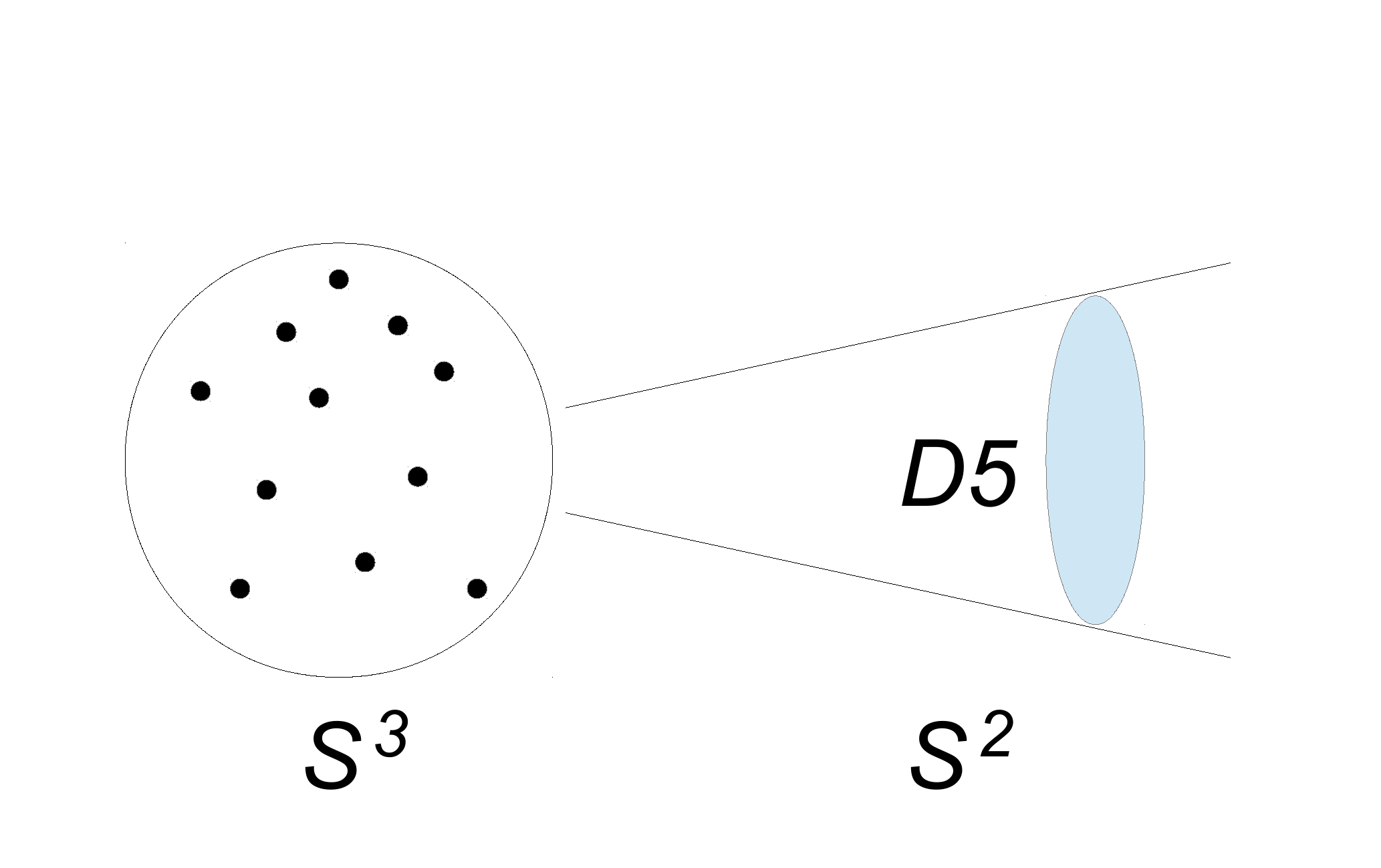}
\end{array}$
\caption{\emph{Left}: localized anti-D3 branes at the north pole of the $S^3$ can polarize into a NS5 brane wrapping a two sphere $S^2 \subset S^3$ and into a D5 brane wrapping the shrinking $S^2$ of the conifold. \emph{Right}: smearing the anti-D3 branes on the $S^3$ wipes out the KPV channel but the D5 channel still survives.}
\label{concept}
\end{center}
\end{figure}

Thus, the healthy instinct when seeing a singular solution is to discard it, unless there is a good physical reason to accept it. For anti-D3 branes in Klebanov-Strassler one would expect that there is such a reason\footnote{We thank H. Verlinde and J. Maldacena for discussions on this point.}:  brane polarization \cite{Myers:1999ps} \`a la Polchinski-Strassler \cite{Polchinski:2000uf}. Indeed, in the probe approximation, the probe anti-D3 branes were found to polarize into NS5 branes that wrap a contractible $S^2$ inside the large $S^3$ at the tip of the conifold \cite{Kachru:2002gs}, as drawn in Figure \ref{concept}, and one might expect that this polarization will continue to happen in the fully backreacted solution. However, to check this directly one would need to construct a solution for multiple anti-D3 branes localized at the north pole of the $S^3$. Unfortunately, constructing non-supersymmetric solutions that depend on two variables is beyond current technology\footnote{Only the supersymmetric KS solution with localized D3 branes is known \cite{Krishnan:2008gx}, \cite{Pufu:2010ie}.}, so the resolution of the singularity via polarization into NS5 branes cannot be directly checked.

However, one can use a less direct route to this result by remembering a very important feature of the Polchinski-Strassler construction: the D3 branes that polarize into NS5 branes wrapping an $S^2$ inside a three-plane can also polarize into D5 branes wrapping an $S^2$ inside an orthogonal plane, and more generally into a $(p,q)$ five-brane wrapping an $S^2$ inside a diagonal three-plane. Hence, if the NS5 polarization channel is present and can cure the anti-brane singularity, so should be the other $(p,q)$ five-brane channels, as well as the D5 channel. The latter channel would correspond to a polarization of the anti-D3 branes localized at the north pole of a large $S^3$ at the tip of KS into a D5 brane wrapping the contractible $S^2$ of the deformed conifold at a finite distance away from the KS tip  (see Figure \ref{concept}). At fist glance this calculation looks as hopeless as the previous one, as it appears to also require the backreacted localized anti-D3 solution. However, things are much better: as shown in \cite{Polchinski:2000uf} and as we will review in Section \ref{TheMeanField}, the polarization potential is independent of the location of the branes that polarize, and hence the potential for the D3's to polarize into D5 branes can be calculated from the smeared near-antibrane solution. The purpose of this paper is to calculate this polarization potential.

We find that this potential has exactly the same type of terms as the polarization potential in Polchinski-Strassler, which confirms the expectation that brane polarization may be the mechanism of choice for resolving this singularity. However,
the coefficients of the terms are not the same as in \cite{Polchinski:2000uf}; in particular, these coefficients depend nontrivially on two parameters that can only be fixed if one knows the full interpolating solution between the IR and the UV. Thus, in general, our potential could have had either SUSY minima (as in \cite{Polchinski:2000uf}), stable non-SUSY minima, metastable minima, or no polarization whatsoever. However, we find that, for \emph{any values} of the unknown parameters, the terms are such that \emph{no} polarization is possible.

Hence, our calculation shows that the singularity of the smeared anti-D3 infrared solution of \cite{Bena:2012bk} \emph{cannot} be resolved by brane polarization, and by the arguments above, that also the localized anti-D3 brane solution will not be resolvable by polarization into D5 branes. Of course, our result does not directly rule out a resolution of the antibrane singularity in KS by polarization into NS5 branes that wrap the $S^2$ inside the $S^3$ at the tip \`a la KPV. However, the fact that D3 brane polarization always happens in multiple channels, and the fact that at least one of this channels is absent, suggests that the KPV polarization channel into NS5 branes might also be absent at full backreaction.

Our calculation further strengthens the evidence that anti D-branes in solutions with charge dissolved in fluxes do not give rise to metastable vacua. Two immediate corollaries follow: the first one is that the dual gauge theories, despite having a intricate structure of supersymmetric vacua \cite{Dymarsky:2005xt}, do not have metastable vacua. The other
is that the mechanism for uplifting AdS vacua with stabilized moduli to dS vacua by adding anti-D3 branes in regions of high warp factor \cite{Kachru:2003aw} will probably not work and will have to be replaced by another uplift mechanism. While there are several other uplift mechanisms in the market (F/D-term uplifting \cite{Saltman:2004sn,Lebedev:2006qq} and Kahler-uplift \cite{Balasubramanian:2005zx,Rummel:2011cd}), none is as generic as anti-D3 uplifting, and it may be possible that these mechanisms will also suffer from similar problems. Hence, it may be necessary to revisit the idea that string theory has a large landscape of dS vacua and to fall back to the old ``non-anthropic'' approach to understanding the physics of our universe.

The paper is organized as follows. In the next section we start with the generalities of Klebanov-Strassler solution. We discuss the IR boundary conditions for anti-D3's smeared over the deformed tip of the conifold and identify the criteria for 3-form flux regularity. In Section \ref{ARegularSolution} we demonstrate that the regular solution does not exist, namely one cannot connect smoothly the anti-D3 region in the IR to the KS solution in the UV. We provide two different proofs: the first one is based on the discussion in~\cite{Bena:2012bk}, while the second is brand-new. We then investigate the behavior of the singular solution in Section \ref{TheSingularSolution}. These results are further used in Section \ref{D5polarization} where we calculate the polarization potential and argue that it has no metastable minima. We also compare the potential to the one of~\cite{Bena:2012tx} corresponding to polarization of anti-D6 branes into D8-branes. We summarize our results in the last section and briefly point towards future directions.
Various useful formulae are relegated to the appendices.

\section{The setup}
\label{TheSetup}

We solved for the full backreaction of anti-D3 branes in the near tip region. In this section we will introduce a useful computational technique and we describe the essential features of regular supersymmetric solutions of the equations of motion, while in the next section we expand the results presented in~\cite{Bena:2012bk}, and we prove that there is no singularity-free solution corresponding to smeared anti-D3 branes at the tip of the deformed conifold.

\subsection{The Papadopoulos-Tseytlin Ansatz}

In \cite{Papadopoulos:2000gj} Papadopoulos and Tseytlin (PT) wrote down the most general form of the warped-conifold type IIB background that preserves the $SU(2) \times SU(2)$ isometry. In this paper we will study the backreaction of (anti)D3 branes \emph{smeared} over the tip of the deformed conifold. This backreaction preserves an additional $\mathbf{Z}_2 \in U(1)_{\rm R}$ symmetry of the Klebanov-Strassler (KS) geometry, otherwise broken in more general PT solutions (like the one describing the full baryonic branch of KS \cite{Butti:2004pk}).

The Ansatz for the solution describing smeared D3 or anti-D3 branes in the KS background is given by:
\begin{eqnarray}
\label{setup}
    d s_{10}^2 &=& e^{2\, A+2\, p-x}\, d s_{1,3}^2 + e^{-6\, p-x}\, \left( d\tau^2 + g_5^2 \right)
                   + e^{x+y}\, \left( g_1^2 + g_2^2 \right) + e^{x-y}\, \left( g_3^2 + g_4^2 \right) \, \label{PTmetric} \\
 C_0 &=& 0 \nonumber \\
 H_3 &=& \frac{1}{2} \left( k - f \right) \, g_5 \wedge \left( g_1 \wedge g_3+ g_2 \wedge g_4 \right)
         + \, d\tau \wedge \left( \dot{f} \, g_1 \wedge g_2 + \dot{k} \, g_3 \wedge g_4 \right)  \nonumber  \\
 F_3 &=& F \, g_1 \wedge g_2 \wedge g_5 + \left( 2 P - F \right) \, g_3\wedge g_4 \wedge g_5
         + \, \dot{F} \, d \tau \wedge \left( g_1 \wedge g_3 + g_2 \wedge g_4 \right) \,  \label{PTfluxes} \\
 F_5 &=& {\cal F}_5 + * {\cal F}_5 \, ,
    \quad {\cal F}_5 = K \, g_1 \wedge g_2 \wedge g_3 \wedge g_4 \wedge g_5 \, , \nonumber
\end{eqnarray}
with
\begin{equation}
\label{K}
   K \equiv  - \frac{\pi}{4} Q + (2P - F) f + k F \, ,
\end{equation}
where all the functions (including the dilaton) depend only on the radial variable $\tau$ and the dot stands for the  $\tau$-derivative. The angular forms $g_i$ are defined in \cite{Klebanov:2000hb}, and are such that $g_1$ and $g_2$ are along the shrinking $S^2$, while $g_{3,4,5}$ are along the $S^3$ of finite size at the tip.
The constant $P$ is proportional to the $5$-brane flux of the KS solution
\begin{equation}
P = \frac14 M \alpha ' \, ,
\end{equation} while $Q$ is the number
of (anti) D3 branes.\footnote{The $\pi/4$ coefficient ensures that $Q$ is indeed the number of D3 branes.
Throughout this paper we will use the conventions of \cite{Papadopoulos:2000gj} apart from the definition of $Q$. In our conventions $Q$ is positive for D3 branes and negative for anti-D3's (and similarly $K$ is negative in KS and positive in anti-KS). These are the same conventions as in \cite{Bena:2011wh}.} In the following we will set $\alpha' =1$.

After integration over all but the radial coordinate, the type IIB supergravity action reduces to an effective \emph{one-dimensional} action for the eight fields
\beq
\phi^a = \left(x, p, A, y, f, k, F, \Phi \right) \ ,
\eeq
whose kinetic term and  potential are:
\begin{eqnarray}
\label{GabV}
G_{a b} (\phi) \phi^a \phi^b &=& - e^{4(p + A)} \, \Big( \dot{x}^2 + \frac{\dot{y}^2}{2} + \frac{\dot{\Phi}^2}{4}
            + 6 \left( \dot{p}^2 - \dot{A}^2 \right) \\
            &&
            + \frac{1}{4} e^{- \Phi - 2 x} \left( e^{-2 y} \dot{f}^2 + e^{2 y} \dot{k}^2 \right)
            + \frac{1}{2} e^{\Phi - 2 x} \dot{F}^2 \Big)  \,     \nonumber\\
V &=& - e^{4(p + A)} \Big( \frac{1}{4} e^{-12 p - 4 x} - e^{-6 p - 2 x} \cosh y + \frac{1}{4} \sinh^2 y  \nonumber\\
  &&  + \frac{1}{16} \left( e^{-\Phi - 2 x} (f - k)^2 +
               2 e^{\Phi - 2 x} \left( e^{-2 y} F^2 + e^{2 y} (2P - F)^2 \right) + 2 e^{-4 x} K^2 \right) \, .\nonumber
     \Big)
\end{eqnarray}
This potential $V$ can, in turn, be obtained from a superpotential $W$:
\begin{equation}
\label{V}
    V = \frac{1}{8} G^{ab} \frac{\partial W}{\partial \phi^a} \frac{\partial W}{\partial \phi^b} \, .
\end{equation}
In fact this equation has two different solutions, and therefore $V$ has {\em two} possible superpotentials:
\begin{equation}
\label{W}
    W^{\pm} = e^{4(p + A)} \left( \cosh y \pm e^{-6 p - 2 x} \pm \frac{1}{2} e^{-2 x} K \right) \, .
\end{equation}
Using either of the two $W$, the supersymmetry conditions can be neatly written as a first-order flow equation
\begin{equation}
\label{firstorderflow}
     G_{a b} \dot{\phi}^b - \frac{1}{2} \frac{ \partial W}{ \partial \phi^a}  = 0\, .
\end{equation}

The presence of the two superpotentials follows directly from the invariance of the type IIB action under the  flip $(C_4, H_3) \to (-C_4, -H_3)$. In our notations this corresponds to the change of sign of $f$, $k$, $Q$ and, as a result, of $K$. The first-order equations following from the two superpotentials impose either an \emph{imaginary self-duality} (ISD) or \emph{imaginary anti-self-duality} (IASD) condition on the complex 3-form, $G_3 \equiv F_3 + i e^{-\Phi}H_3$. As the subscript suggests, in our conventions, the supersymmetric solution derived from $W^+$ is the Klebanov-Strassler background with ISD 3-form, while $W^-$ leads to the anti-Klebanov-Strassler solution with IASD fluxes. The two solutions preserve different supersymmetries. Consequently the supersymmetric KS solution can also include arbitrary number of the mobile D3-branes, $Q>0$, but no anti-D3 branes, $Q<0$; and vice versa for the supersymmetric anti-KS background.

\subsection{The KS and anti-KS solutions}

Before proceeding it is worth mentioning how the eight integration constants of the eight first-order superpotential equations~\eqref{firstorderflow} (with $W=W^+$) are fixed in the KS solution with $Q$ smeared mobile D3 branes ($Q>0$):
\begin{itemize}
  \item The zero-energy condition of the effective Lagrangian fixes the $\tau$-redefinition gauge
        freedom and is automatically solved, but the constant shift
        $\tau \to \tau + \tau_0$ still remains unfixed, and so $\tau_0$ appears as a ``trivial" integration constant.
  \item The conifold deformation parameter $\epsilon$ and the constant dilaton $e^{\phi_0}$ give
        two other free parameters.
  \item An additional parameter renders the conifold metric singular in the IR \cite{Candelas:1989js} and has to be discarded.
  \item The three equations for the flux functions $f$, $k$ and $F$ appear to have three free parameters \cite{Kuperstein:2003yt}. The first corresponds to an IR singular $(2,1)$ complex 3-form $G_3 \equiv F_3 + i e^{-\phi} H_3$, the second gives a $(0,3)$ form which is singular in the UV\footnote{Importantly the singular $(2,1)$ form is supersymmetric exactly as the 3-form of the KS solution, while  the $(0,3)$ form breaks SUSY \cite{Grana:2000jj,Gubser:2000vg}.}, and the third is related to the $B$-field gauge transformation $(f,k) \to (f + c, k + c)$, which is just a shift of the D3 brane charge and can be absorbed in the redefinition of $Q$.
  \item The warp function $h \sim e^{2 x - 4(p + A)}$ can only be determined up to a constant,
        which is fixed requiring that $h(\tau)$ vanishes at infinity:
       \begin{equation}
        \label{h}
          h(\tau) =  \, \int_{\tau}^{\infty} d \bar{\tau}
              \frac{
                 \left( 4 \pi Q  +
              32 g_s P^2 \, \left( \bar{\tau} \coth (\bar{\tau}) - 1 \right) \left( \sinh (\bar{\tau}) \right)^{-2}
                         \left( \frac{1}{2} \sinh(2 \bar{\tau}) - \bar{\tau} \right)
                 \right)  }{\left( \frac{1}{2} \sinh(2 \bar{\tau}) - \bar{\tau} \right)^{2/3}}
                          \, .
       \end{equation}
       It is important to stress here that for the anti-KS solution with anti-D3's ($Q<0$) one has to put $\left\vert{Q}\right\vert$ instead of $Q$, since otherwise $h(\tau)$ is negative for small $\tau$. This is contrary to the flux $K(\tau)$ which flips sign once we go from the KS to the anti-KS solution.
\end{itemize}
We have relegated the remaining functions appearing in the Klebanov-Strassler solution to Appendix \ref{AppendixKS}. As has been explained above, the anti-KS solution can be easily found by flipping the sign of the functions  $f$ and $k$. Notice that the $(2,1)$ and the $(0,3)$ 3-forms will be now $(1,2)$ and $(3,0)$. The remaining functions are exactly the same for the solutions derived from $W^+$ and $W^-$.


\subsection{The first-order formalism}

In order to solve for the anti-D3 backreaction we will need to solve the full set of second-order equations of motion coming from~\eqref{GabV}. We now introduce a computational technique that will be extremely useful: the idea is to recast the eight second-order EOMs for the scalars $\phi^a$ as a set of sixteen coupled first-order equations by introducing conjugate momenta $\xi_a$, defined as
\begin{equation}
\label{xi}
   \xi_a=  G_{a b} \dot{\phi}^b - \frac{1}{2} \frac{ \partial W}{ \partial \phi^a}\, .
\end{equation}

Since we have two superpotentials that govern the system,  $W^+$ and $W^-$, we can introduce {\em two} sets of conjugate modes, denoted by $\xi_a^+$ and $\xi^{-}_a$ respectively. With this notation the supersymmetric KS first-order flow equations (with ISD fluxes) are simply $\xi_a^+ = 0$, while the first-order equations corresponding to supersymmetric anti-KS solutions (with IASD fluxes) are $\xi^{-}_a = 0$.
It is easy to verify that solutions of these eight first-order equations solve also the full set of EOM. Indeed, by plugging the definition~\eqref{xi} into the second order EOMs we obtain:
\begin{equation}
\label{dxi}
    \dot{\xi}_a = - \frac{1}{2} \left[ \frac{\partial G^{b c}}{\partial \phi^a} \xi_b \xi_c
                                   + \frac{\partial G^{b c}}{\partial \phi^a} \frac{\partial W}{\partial \phi^b} \xi_c
                                   + G^{b c} \frac{\partial^2 W}{\partial {\phi^a} \partial {\phi^b}} \xi_c
                                   \right] \, ,
\end{equation}
which is indeed trivially solved by putting all of $\xi_a$'s to zero.

Replacing the second-order EOMs for the eight fields $\phi_a$ by sixteen first-order ones, equations~\eqref{xi} and~\eqref{dxi}, proves very efficient when studying supersymmetry breaking perturbatively \cite{Borokhov:2002fm,Bena:2009xk}, and turns out to be extremely useful for our purpose as well. As we will review in the next section, it was shown in \cite{Bena:2012bk}  that without introducing singular fluxes it is not possible to interpolate between the ISD Klebanov-Strassler solution in the UV and the anti-D3 branes ($Q<0$) boundary conditions in the IR. The regularity conditions on the fields near the anti-branes determine almost uniquely the leading-order behavior of the fields $\xi_a$'s derived from $W^-$, which in turn appears to be incompatible with the equations (\ref{dxi}).
Moreover, we will provide a new ``topological" argument leading to the same conclusion but using instead the $\xi_a$ functions derived from $W^+$. Since we will make an extensive use of both functions $\xi$,
 the following should useful to keep track of the notation:
\begin{eqnarray}
W^+ = W_{\rm KS} \ , \quad &{\rm BPS \, solution} \ : \,  \xi^+=0 \ , \quad G_3  \, {\rm ISD}  \ , \quad F_{D3}=0 \nonumber \\
W^- = W_{\rm AKS} \ , \quad &{\rm BPS \, solution} \ :  \, \xi^-=0 \ , \quad G_3 \,  {\rm IASD} \ , \quad F_{\overline{D3}}=0 \nonumber
\end{eqnarray}
where in the last equality we have added the force on probe D3 and anti-D3 branes.

The explicit form of (\ref{xi}) for the conjugate momenta is:
\begin{align}
  \xi_1^{\pm}     &=  - e^{4(p + A)} \left( \dot{x} - 2 \dot{p} - 2 \dot{A} \mp \frac{1}{2} e^{-2 x} K  \right) \nonumber \\
  \xi_2^{\pm}    &=  -  e^{4(p + A)} \left( \dot{x} + \dot{p} - 2 \dot{A} + \cosh y - \frac{1}{2} e^{- 6 p -2 x} \right)  \nonumber \\
  \xi_3^{\pm}     &=  - 6 e^{4(p + A)} \left( \dot{p} + \dot{A} - \frac{1}{2} e^{-6 p - 2 x} \right) \nonumber
\end{align}
\begin{align}
  \xi_y^{\pm}     &=  - \frac{1}{2} e^{4(p + A)} \left( \dot{y} + \sinh y \right) \nonumber \\
  \xi_{\Phi}^{\pm}  &=  - \frac{1}{4} e^{4(p + A)} \dot{\phi}  \nonumber \\
  \xi_f^{\pm}     &=  - \frac{1}{4} e^{-2 x + 4(p + A)} \left( e^{- \Phi - 2 y} \dot{f} \pm (2 P - F) \right)  \nonumber \\
  \xi_k^{\pm}     &=  - \frac{1}{4} e^{-2 x + 4(p + A)} \left( e^{- \Phi + 2 y} \dot{k} \pm F \right)  \nonumber \\
  \xi_F^{\pm}     &=  - \frac{1}{2} e^{-2 x + 4(p + A)} \left( e^{\Phi} \dot{F} \pm \frac{1}{2} \left(k - f \right) \right) \, ,\label{xi-eq}
\end{align}
where
\begin{equation}
 \xi_1^{\pm} \equiv \xi_x^{\pm} - \frac{\xi_p^{\pm}}{3} + \frac{\xi_A^{\pm}}{3} \, , \,
 \xi_2^{\pm} \equiv \xi_x^{\pm} + \frac{\xi_p^{\pm}}{6} + \frac{\xi_A^{\pm}}{3} \, , \,
 \xi_3^{\pm} \equiv \xi_p^{\pm} - \xi_A^{\pm} \, .
\end{equation}
The $\xi_1$ redefinition will prove to be especially convenient, since we can show that this mode has a very clear physical meaning: it parameterizes the force felt by D3 branes probing a given solution. Indeed, the force on probe D3 and anti-D3 branes only depends on $\xi_1$ and no other $\xi_a$:
\begin{equation}\label{forceD3}
F_{D3} = - 2 e^{-2x} \xi_1^+ \ , \quad F_{\overline{D3}} = - 2 e^{-2x} \xi^-_1\, .
\end{equation}
As expected, adding a probe D3 brane to a solution derived from the superpotential $W^+$ (with ISD fluxes) does not break supersymmetry, and hence the force on probe D3 branes, $F_{D3}$, vanishes. Analogously, an anti-D3 brane in the anti-KS solution does not break supersymmetry and therefore feels no force. In a general non-supersymmetric solution, such as the singular anti-D3 in KS  backreacted solution that we analyze in Section \ref{TheSingularSolution}, both forces are nonzero.

\section{A regular solution does not exist}
\label{ARegularSolution}

In this section we review and expand the results of~\cite{Bena:2012bk}. We will prove that there is no IR-regular solution
with smeared anti-D3 branes ($Q<0$) at the tip of the conifold and with KS asymptotics in the UV.
Indeed, starting with a singularity-free anti-brane solution in the IR, one necessarily ends up with an anti-KS solution in the UV.
Moreover, we will prove that the only regular solution with $|Q|$ anti-D3 branes is \emph{exactly} the anti-KS version of the solution with $Q$ mobile anti-branes we described in the previous subsection.

\subsection{Regular boundary conditions for anti-D3 branes}
\label{subsec:regularIR}

In order to prove our statement, we need to understand first the IR boundary conditions corresponding to the presence of smeared anti-D3 branes at the tip of the KS geometry. We will also impose regularity of the 3-form fluxes. These conditions, which we will call IR regularity conditions, are the following:
\begin{itemize}
  \item The dilaton is finite at $\tau=0$.
  \item the $6d$ conifold metric has the tip structure of the KS solution:
        the 2-sphere (the $g_1^2+g_2^2$ part of the $6d$ metric) shrinks smoothly at $\tau=0$ and the 3-sphere
        (the $g_3^2 + g_4^2 + \frac{1}{2} g_5^2$ term) has finite size. The former condition is equivalent to $2 e^{-6 p - x} \approx e^{x - y}$ near $\tau=0$, and the latter requires $\tau^2 e^{-6 p -x} \approx 2 e^{x + y}$. All in all, we find that
        \begin{equation}
        \label{IR1}
            e^{6 p + 2 x} = \tau + \mathcal{O} (\tau^2) \, , \qquad  e^y = \frac{\tau}{2} + \mathcal{O} (\tau^2) \, .
        \end{equation}
  \item The warp factor comes from $|Q|$ anti-D3 branes smeared on the 3-sphere,
        and hence goes like  \mbox{$ h(\tau) \sim |Q|/\tau$}. In our notation it amounts to demanding that both
        $e^{12 p + 2 x}$ and $e^{4(p+A) - 2x}$ go like $\tau$. The precise proportionality coefficients, though, cannot be fixed in this approach. Instead, one coefficient can be eliminated by a proper rescaling of the $4d$ space-time coordinates, while the other is a free parameter that measures the size of the non-shrinking 3-sphere (the conifold deformation parameter $\epsilon$). We will use the $4d$ rescaling to match the expansion of $e^{6 p + x}$ to the supersymmetric solution (see (\ref{h}) and (\ref{KS2})):
        \begin{equation}
        \label{IR2}
            e^{12 p + 2 x} = \frac{4}{\pi Q} \cdot \tau + \mathcal{O} (\tau^2) \, , \qquad
            e^{4(p+A) - 2x} = c_0 \frac{4}{\pi Q} \cdot \tau + \mathcal{O} (\tau^2) \, .
        \end{equation}
        For the KS solution one finds $c_0 = 2^{-10/3} 3^{-2/3} \epsilon_0^{8/3}$.
  \item There is no singularity in the three-form fluxes; their energy densities, $H_3^2$ and $F_3^2$,
        do not diverge at  $\tau=0$. From (\ref{setup}) we obtain that
        \begin{align}
        \label{SquaredH3F3}
            \left\vert F_3 \right\vert^2 & = F_{\mu \nu \rho}F^{\mu \nu\rho} =
                 6 e^{6 p - x } \left( e^{2y} (2 P - F)^2 + e^{-2y} F^2 + 2 \dot{F}^2 \right)  \nonumber \\
            \left\vert H_3 \right\vert^2 & = H_{\mu \nu \rho}H^{\mu \nu\rho} =
                 6 e^{6 p - x } \left( e^{-2y} \dot{f}^2 + e^{2y} \dot{k}^2 + \frac{1}{2} (k-f)^2 \right) \, .
        \end{align}
        Hence, using (\ref{IR1}) and (\ref{IR2}) the Taylor expansions of the functions $f$, $k$ and  $F$
        start from $\tau^3$, $\tau$ and $\tau^2$ terms respectively,
        exactly like in the KS background (see Appendix \ref{AppendixKS}).
        To be more precise, in a solution with branes at the tip, the functions $f$, $k$ and $F$
        can also start with non-integer powers ($\tau^{9/4}$, $\tau^{1/4}$ and $\tau^{5/4}$),
        but it is not hard to show that the logarithmic terms in $x$, $p$, $A$ and $y$ imply that the IR expansion of the
        solution proceeds only with integer powers of $\tau$. In either situation the expansion of
        $K$ starts with a constant $Q$ term.
\end{itemize}
Let us summarize the leading IR terms in the expansion of the metric functions:
\begin{align}
\label{IRexpansion}
&   e^{\Phi} = e^{\Phi_0} + \mathcal{O}(\tau) \, , \qquad
    e^{2x} = \frac{\pi Q}{4} \cdot \tau + \mathcal{O} (\tau^2) \, , \qquad
    e^y = \frac{\tau}{2} + \mathcal{O} (\tau^2) \, , \, , \nonumber \\
&   \qquad \qquad
    e^{6 p} = \frac{4}{\pi Q} + \mathcal{O} (\tau) \, , \qquad
    e^{6 A} = c_0^\frac{3}{2} \frac{\pi Q}{4} \cdot \tau^3 + \mathcal{O}(\tau^4) \, , \\
&
    f  = \mathcal{O}(\tau^3) \, , \qquad k = \mathcal{O}(\tau) \, , \qquad F  = \mathcal{O}(\tau^2) \, , \qquad
    K = - \frac{\pi Q}{4} + \mathcal{O} (\tau^3)\, . \nonumber
\end{align}
Even if we arrived at the boundary conditions~\eqref{IRexpansion} by physical arguments, one my wonder whether these are the most general conditions we can impose. We checked that if we start by allowing a general Taylor expansion for the functions $x$, $y$, $p$ and $A$, the equations of motion imply precisely the behavior summarized in~\eqref{IRexpansion}.

For our proof that this IR behavior does not glue to a solution with ISD fluxes in the UV, it will be essential to determine the leading-order behavior of the conjugate modes $\xi_a^+$'s and $\xi^{-}_a$'s (defined in~\eqref{xi-eq}) in the IR. Let us denote by $n^+_a$ and $n^-_a$ the lowest possible leading orders of these two functions respectively.
We find that the boundary conditions~\eqref{IRexpansion} imply:
\begin{eqnarray}
\label{powers}
   \left( n_1^+, n_2^+, n_3^+, n_y^+, n_f^+, n_k^+, n_F^+, n_\Phi^+ \right) &=&  \left( 1, 2, 2, 2, 1, 3, 2, 2 \right) \nonumber\\
   \left( n^-_1, n^-_2, n^-_3, n^-_y, n^-_f, n^-_k, n^-_F, n^-_\Phi \right) &=&
            \left( 2, 2, 2, 2, 1, 3, 2, 2 \right) \, .
\end{eqnarray}
The only difference between the two sets is in $n_1^+$ and $n^-_1$. Indeed, from (\ref{IRexpansion}) one sees that
the leading (logarithmic) terms cancel out in the parenthesis of $\xi^{-}_1$, eq. (\ref{xi-eq}), and sum up for $\xi_1^+$. Similar cancelations happen also for $\xi_2^+$, $\xi_3^+$, $\xi_y^+$ and their $\xi^{-}_a$ counterparts. However, for the 3-form $\xi_a$'s we cannot argue for such a cancelation neither for  $\xi_f^+$, $\xi_k^+$ and $\xi_F^+$
nor for $\xi^{-}_f$, $\xi^{-}_k$ and $\xi^{-}_F$. This is since we have no control
over the coefficients of the leading terms in the expansions of $f$, $k$ and $F$.

It is important to stress again that in arriving at (\ref{powers}) we have not imposed neither the ISD nor the IASD flux condition. Instead, we insisted on having a regular 3-form flux in the IR, with all other components of the solution being that of a smeared D3-brane solution: $1/\tau$ behavior of the warp function, constant 5-form flux proportional to $Q$, plus a constant dilaton.


Finally, we should also briefly mention the UV boundary conditions, although their details will not be used in our discussion. In general we must insist on KS (and not anti-KS) asymptotic with some \emph{normalizable} UV modes turned on. Having only normalizable modes in the UV should be essential for the construction, since the new solution must describe a new vacuum in the same theory.
Since in the UV region the non-supersymmetric solution should be just a small perturbation of the KS solution, one can use the linearized version of the equations of motion. A careful analysis reveals (see, for example, \cite{Bena:2011wh}) that $\xi_f^+(\tau)$ and $\xi_k^+(\tau)$ approach the same non-zero constant value at large $\tau$, while all the other functions $\xi_a^+(\tau)$ vanish.


We would like to demonstrate now that one cannot meet both the IR and the UV boundary conditions advocated in the previous section. We will do it in two different ways. We will find that the only possible solution is $\xi^{-}_a=0$ for all $a$'s, meaning that one has anti KS solution not only in the IR, but also all the way to the UV. Hence, any solutions with anti-D3 branes in the infrared must necessarily have
singular three-form fluxes. This result is in agreement with the linearized analysis of~\cite{Bena:2009xk,Bena:2011wh}, where the equations of motion~\eqref{xi}-\eqref{dxi} were solved perturbatively in the number of antibranes.  Similar results were obtained for other types of anti-branes in background with opposite charge dissolved in fluxes~\cite{Bena:2010gs,Massai:2011vi,Giecold:2011gw,Blaback:2010sj,Blaback:2011nz,Blaback:2011pn}.

We will provide two proofs of this claim.
First, we will argue that the IR conditions (\ref{powers}) are in odds with the $\xi^{-}$'s equations of motion.  This analysis has been carried out originally in \cite{Bena:2012bk}, where it was referred to as  the ``IR obstruction". Second, we will present a new ``global" argument which is also based on the $\xi^{-}$'s equations of motion, but does not employ the Taylor expansion of these functions.

\subsection{The first proof}

Our immediate goal is to show that when solving the equations (\ref{dxi}) for $\xi^{-}_1$, $\xi^{-}_f$, $\xi^{-}_k$ and $\xi^{-}_F$ in the IR (small $\tau$) and imposing the IR regularity conditions, one finds only trivial solutions for these functions. This essentially means that the IASD conditions $\xi^{-}_f, \xi^{-}_k, \xi^{-}_F = 0$ will be satisfied all the way to the UV and not only at $\tau=0$.

The equations we need are:
\begin{equation}
\label{dx1-eq}
    \dot{\xi}^{-}_1 + K e^{-2 x} \xi^{-}_1  =
      4 e^{2 x - 4(p + A)}
        \left[ e^{\Phi+2 y}  (\xi^{-}_f)^2 + e^{\Phi - 2 y} (\xi^{-}_k)^2
        + \frac{1}{2} e^{-\Phi} (\xi^{-}_F)^2 \right]
\end{equation}
and
\begin{eqnarray}
\label{dxfkF-eq}
    \dot{\xi}^{-}_f &=& \frac{1}{2} e^{-2 x} (2 P - F) \xi^{-}_1 + \frac{1}{2} e^{-\Phi} \xi^{-}_F    \nonumber \\
    \dot{\xi}^{-}_k &=& \frac{1}{2} e^{-2 x} F \xi^{-}_1 - \frac{1}{2} e^{-\Phi} \xi^{-}_F   \\
    \dot{\xi}^{-}_F &=& \frac{1}{2} e^{-2 x} (k - f) \xi^{-}_1 +
                    e^{\Phi} \left( e^{2 y} \xi^{-}_f - e^{-2 y} \xi^{-}_k \right)   \, .   \nonumber
\end{eqnarray}
We give the remaining four $\dot{\xi}^{-}_a$ equations in Appendix \ref{DotTildeXi}.

Remember that if the fluxes are regular, the IR expansions of $f(\tau)$, $k(\tau)$ and $F(\tau)$ can only  start from
$\tau^3$, $\tau$ and $\tau^2$ respectively (see the discussion around (\ref{IRexpansion})). As we have pointed out earlier, lower but non-integer powers are not ruled out. One can easily check, though, that our proof still goes through even in this situation.

Let us denote by $n$ the lowest power in the Taylor expansion of $\xi^{-}_1$, i.e.
$\xi^-_1=a_1 \tau^n + \ldots \,$ ,
We already know from (\ref{powers}) that
$n \geqslant 2$. Together with (\ref{IRexpansion}),  equation (\ref{dx1-eq}) implies that the leading terms in the expansions of $\xi^{-}_f$, $\xi^{-}_k$ and $\xi^{-}_F$ are
\begin{equation}
    \xi^{-}_f = a_f \, \tau^{(n-2)/2} + \ldots \, , \quad
    \xi^{-}_k = a_k \, \tau^{(n+2)/2} + \ldots \, , \quad
    \xi^{-}_F = a_F \, \tau^{n/2} + \ldots \, ,
\end{equation}
Note that an additional comparison with (\ref{powers}) shows that actually for a regular solution $n \geqslant 4$.
Moreover, since all the terms on the right hand side of (\ref{dx1-eq}) are non-negative and
$K e^{-2 x} = \tau^{-1} + \ldots$, at least one of the constants
$a_f$, $a_k$ and $a_F$ has to be non-zero. Next, plugging these expansions into the last two equations of (\ref{dxfkF-eq}) we see that for $n \geqslant 4$, the terms involving $\xi^{-}_1$ and  $\xi^{-}_f$ disappear from the leading-order expansions of all these equations.
A simple calculation then reveals that (\ref{dxfkF-eq}) has only two possible solutions, $\xi^{-}_F \sim \tau$ or $\xi^{-}_F \sim \tau^{-2}$, and both yield a singular 3-form flux.\footnote{We will come back to the $\xi^{-}_F \sim \tau$ singular solution in the next section.}
Thus we have to put $a_k, a_F = 0$, in which case the first equation in (\ref{dxfkF-eq}) gives $n=-2$, and so we arrive at a contradiction.

We observe, therefore, that with regular boundary conditions at $\tau=0$, the equations (\ref{dx1-eq}) and (\ref{dxfkF-eq}) have only the trivial solution $\xi^{-}_1=\xi^{-}_f=\xi^{-}_k=\xi^{-}_F=0$. This means that we obtain an IASD solution all the way from the IR to the UV. In other words, one cannot ``glue" the solution near the smeared anti D3-branes to the KS solution, since the latter has
an ISD 3-form.

Importantly, with a bit of an effort we can demonstrate that the anti-KS geometry with mobile anti-D3's at the tip is the \emph{only} regular solution of the remaining equations of motion. In other words, there is no non-singular solution with anti-D3 branes in the IR and anti-KS asymptotics in the UV. To do this we have to prove that all the remaining $\xi^{-}$ functions identically vanish, exactly as $\xi^{-}_1$, $\xi^{-}_f$, $\xi^{-}_k$ and $\xi^{-}_F$.

Plugging $\xi^{-}_{1,f,k,F}=0$ into (\ref{DotTildeXi23yPhi}) we find that $\xi^{-}_\Phi=0$
(otherwise the dilaton diverges), while the remaining functions satisfy:
            \begin{eqnarray}
            \label{xi23y-eq}
                \dot{\xi}^{-}_2  &=& 3 e^{-6 p - 2 x}  \xi^{-}_2 - e^{- 4(p + A)}
                    \left( \frac{2}{3} \xi^{-}_2 \xi^{-}_3 - \frac{1}{18} (\xi^{-}_3)^2 + 2 (\xi^{-}_y)^2 \right)
                \nonumber \\
                \dot{\xi}^{-}_3  &=& 6 e^{-6 p - 2 x} \xi^{-}_2
                \nonumber \\
                \dot{\xi}^{-}_y  &=& \cosh y \, \xi^{-}_y + {1\over 3} \sinh y \, \xi^{-}_3 \, .
            \end{eqnarray}
In the (anti) KS solution $e^{-4 (p + A)}$ goes to zero as $e^{-4 \tau/3}$ for large $\tau$, while $e^{-6 p -2 x}$ asymptotes to $2/3$.
From the first two equations we find that $\dot{\xi}^{-}_2 \approx 2 \xi^{-}_2$.
The functions $\xi^{-}_2$, $\xi^{-}_3$ and $\xi^{-}_y$, exactly like the functions $\xi_2^+$, $\xi_3^+$ and $\xi_y^+$,
have to vanish at infinity both for KS and anti KS solutions.
So we have to put $\xi^{-}_2=0$.
This in turn implies that $\xi^{-}_3=\xi^{-}_y=0$.

We have shown that a regular solution with anti-D3 branes in the IR remains anti-KS all the way to
the UV using the conjugate variables $\xi$. But actually, the most straightforward way to see it is to solve the second-order $\phi_a$ equations directly in powers of $\tau$ subject to the regularity conditions (\ref{IRexpansion}). We found that to order $\tau^{10}$ the space of solutions is parameterized by three
independent parameters, none of which breaks the IASD condition confirming that $\xi^{-}_f$, $\xi^{-}_k$ and $\xi^{-}_F$ are indeed zero for the IR regular solution. Furthermore, one parameter leads to
\begin{equation}
    \xi^{-}_2 = 3 c \tau^3 + \ldots, \qquad
    \xi^{-}_3 = 6 c \tau^3 + \ldots, \qquad
    \xi^{-}_y = -c \tau^3 + \ldots, \, .
\end{equation}
This is consistent with (\ref{xi23y-eq}) and, as we already know, produces a UV divergent solution. The remaining two parameters
correspond to two UV-singular solutions of the supersymmetric $\xi^{-}_a=0$ equations that we have already mentioned in the previous section. The first introduces the $(0,3)$ complex 3-form and the second shifts the warp function.

To sum up, IR regularity of the 3-form fluxes implies that all of the $\xi^{-}_a$'s identically vanish. The integration constants emerging from the $\xi^{-}_a=0$ equations are then fixed by the UV regularity and we end up with the \emph{anti} KS background with $\left\vert Q \right\vert$ mobile \emph{anti} D3 branes.

\subsection{The second proof}

We can also present a ``global" argument why the  functions $\xi^{-}_1$, $\xi^{-}_f$, $\xi^{-}_k$ and $\xi^{-}_F$ have to vanish in a regular solution, without focusing on their Taylor expansions. The proof for the remaining four functions proceeds precisely as above.

Our key observation is that the flux functions $f(\tau)$, $k(\tau)$ and $F(\tau)$ appear only in equations (\ref{dx1-eq}) and
(\ref{dxfkF-eq}). None of the remaining $\dot{\xi^{-}}_a$ equations has any flux function in it. Next, the equations in
(\ref{dxfkF-eq}) might be derived from the following \emph{reduced} Lagrangian:
\begin{equation}
\label{Lreduced}
    \mathcal{L}_{\rm fluxes} = 4 e^{2 x - 4(p + A)}
        \left[e^{\Phi+2 y}  (\xi^{-}_f)^2 + e^{\Phi- 2 y} (\xi^{-}_k)^2
        + \frac{1}{2} e^{-\Phi} (\xi^{-}_F)^2 \right]
        + e^{- 4(p + A)} (\xi^{-}_1)^2  \, .
\end{equation}
Recall that the $\xi^{-}$'s are first order in the derivatives of $\phi$'s and so the Lagrangian is of second order in $\tau$-derivatives, as it should be.
It differs from the second and the fourth lines of (\ref{GabV}) only by total derivative terms. Written this way, however, $\mathcal{L}_{\rm fluxes}$ has a remarkable property: it is strictly non-negative and vanishes only if all the functions  $\xi^{-}_1$, $\xi^{-}_f$, $\xi^{-}_k$ and $\xi^{-}_F$ are zero.

Again, we treat $\mathcal{L}_{\rm fluxes}$ as the effective Lagrangian \emph{only} for the  fields $f(\tau)$, $k(\tau)$ and $F(\tau)$. In particular, it means that the first three terms in (\ref{Lreduced}) are kinetic terms, while the last one is a potential term.
We assume now that one first solves (\ref{dxfkF-eq}) for these three fields and for arbitrary $x$, $y$, $p$, $A$, $\Phi$ (but with the proper boundary conditions ensuring regularity of the metric), and then substitutes the result into the remaining five EOM.

Since $\mathcal{L}_{\rm fluxes}$ is bounded from below, in other words has a global minimum for
\begin{equation}
\label{theglobalminimum}
    \xi^{-}_f(\tau), \quad \xi^{-}_k(\tau), \quad \xi^{-}_F(\tau), \quad \xi^{-}_1(\tau) \quad = \quad 0 \, ,
\end{equation}
one may wonder whether this \emph{trivial} IASD solution is, in fact, the \emph{unique} solution of the EOM (\ref{dxfkF-eq}). The answer depends on the boundary conditions for $f(\tau)$, $k(\tau)$ and $F(\tau)$. If these are \emph{incompatible} with (\ref{theglobalminimum}), the final solution will be more complicated. If on the other hand, the regular boundary conditions we imposed on the 3-form flux are compatible with the trivial IASD solution, then the latter will also be the only possible solution.


For our Lagrangian (\ref{Lreduced}) the fields $\xi^{-}_f$, $\xi^{-}_k$ and $\xi^{-}_F$ are the conjugate momenta of the fields $f$, $k$ and $F$ respectively. In general, one may impose boundary conditions either on these fields or on their conjugate momentum, in the IR or/and in the UV.

The regularity requirement we considered in the previous sections, however, constrains all the three flux functions and their conjugate momenta in the IR. Indeed, we saw that both $(f, k, F)$ and $( \xi^{-}_f, \xi^{-}_k, \xi^{-}_F )$ have to vanish at $\tau=0$ for a regular solution. Furthermore, $\xi^{-}_1=0$ in the IR, therefore the IR boundary conditions following from the regularity are consistent with the trivial solution (\ref{dxfkF-eq}). Thus we see that requiring regularity in the IR forces upon us the anti-KS solution.


This proof, though, has to be taken with a grain of salt, since the EOM for the flux fields are strictly speaking singular at $\tau=0$, and so we cannot rule out completely the possibility that there are two different solutions of (\ref{dxfkF-eq}) subject to the same boundary conditions. One can promptly make our proof more rigid by listing Taylor $\tau^n$-expansions of all six possible solutions in the IR and verifying that only one of them, the IASD, is not at odds with (\ref{theglobalminimum}). However, the main goal of this subsection is to prepare the ground for the localized case discussion, where the power counting method of the first proof will be most likely unavailable making a ``topological" argument we presented here a more efficient tool.

\section{The singular anti-D3 solution}
\label{TheSingularSolution}

In the previous section we proved that by imposing the regular IR boundary conditions summarized in~\eqref{IRexpansion}, it is not possible to find a supersymmetry-breaking solution (except the one that we have mentioned before, corresponding to ISD fluxes with a $(0,3)$ component, which diverges in the UV). Thus, the regular IR boundary conditions are incompatible with the presence of anti-D3 branes in the infrared. One can try to construct a singular solution describing the backreaction of these anti-D3 branes by relaxing the assumptions we made in the previous section, and considering a more general expansion for the fields. In this section we therefore analyze the equations of motion dropping the assumption of regularity in the three-form fluxes discussed in Section~\ref{subsec:regularIR}.

Let us start by noticing that even in a solution with singular 3-forms, all $\xi^{-}_a$'s, but $\xi^{-}_f$, $\xi^{-}_k$ and $\xi^{-}_F$, have the same leading term powers at small $\tau$ as for any regular solution, see (\ref{powers}).
In particular, we still have $\xi^{-}_1 \sim \tau^2$, since otherwise the solution will not describe \emph{anti} D3's at the tip of the conifold. At the same time, $\xi^{-}_f$, $\xi^{-}_k$ and $\xi^{-}_F$ will now start with lower powers of $\tau$.
Remarkably, equation (\ref{dx1-eq}) suffices to determine this behavior. Indeed, since the left hand side is still of order $\tau$ exactly like in the regular case, and the right hand side is a sum of positive terms, we see that:
\begin{equation}
\label{bfkF}
    \xi^{-}_f = b_f + \mathcal{O} (\tau) \, , \quad
    \xi^{-}_k = b_k \, \tau^2 + \mathcal{O} (\tau^3) \, , \quad
    \xi^{-}_F = b_F \, \tau + \mathcal{O} (\tau^2) \,  .
\end{equation}
 In deriving this result we used the first two lines of (\ref{IRexpansion}). The expansions of the original flux functions are:
\begin{equation}
\label{SINGULARfkF}
    f = - \frac{\pi Q}{8 c_0}  e^{\Phi_0} b_f  \, \tau^2 + \mathcal{O} (\tau^3) \, , \quad
    k = \frac{\pi Q}{c_0} \left( b_F + 2 e^{\Phi_0} b_k \right) + \mathcal{O} (\tau) \, , \quad
    F = \frac{\pi Q}{c_0} b_k  \, \tau + \mathcal{O} (\tau^2) \, ,
\end{equation}
where in going from (\ref{bfkF}) to $f(\tau)$, $k(\tau)$ and $F(\tau)$ we have eliminated two additional solutions (see the end of the previous section):  the first one is the ``very'' singular $(1,2)$ solution with $k \sim \tau^{-2}$ and $F \sim \tau^{-1}$ which we will not consider, and the second corresponds to the gauge transformation $(f,k) \to (f + c, k + c)$ we mentioned earlier. We fix the gauge freedom by requiring that $f(\tau)$ vanishes at $\tau=0$.

It seems that, all in all, we have a singular solution in the UV parameterized by three independent parameters $b_f$, $b_k$ and $b_F$. However, only two parameters are independent, since both the $\dot{\xi}^{-}_k$ and the $\dot{\xi}^{-}_F$ equations in (\ref{dxfkF-eq}) imply that
\beq
b_F = - 4 e^{\Phi_0} b_k \ .
\eeq

Thus we have (at least) a two-dimensional space of singular solutions in the IR. At the same time, by gluing the solution to the UV we expect to arrive at a unique solution for the entire range of $\tau$ that depends on two parameters $Q$ and $P$. The UV regularity will then impose an additional constraint on $b_f$ and $b_k$ (as well as on all the other ``free" IR parameters like the dilaton), so that one will have to switch on both these modes in order to avoid a divergent UV solution.
In fact, the perturbative solution constructed in \cite{Bena:2011wh}  at linear order in $Q/P$ has exactly this singularity structure (\ref{SINGULARfkF}) with $b_f = - 12 b_k\sim  0.02 \,\epsilon_0^{8/3}  \, P^{-2}$.
However, for the full solution this result is expected to change.

We see now that the singular solution will necessarily have a non-zero $\xi^{-}_f$ at $\tau=0$. In this case the arguments from the end of the previous section do not apply and, as a result, there is no ``global minimum" obstruction for the singular solution.

\begin{figure}[t]
\begin{center}
$\begin{array}{cc}
\includegraphics[trim = 0mm 120mm 0mm 15mm, scale=0.3]{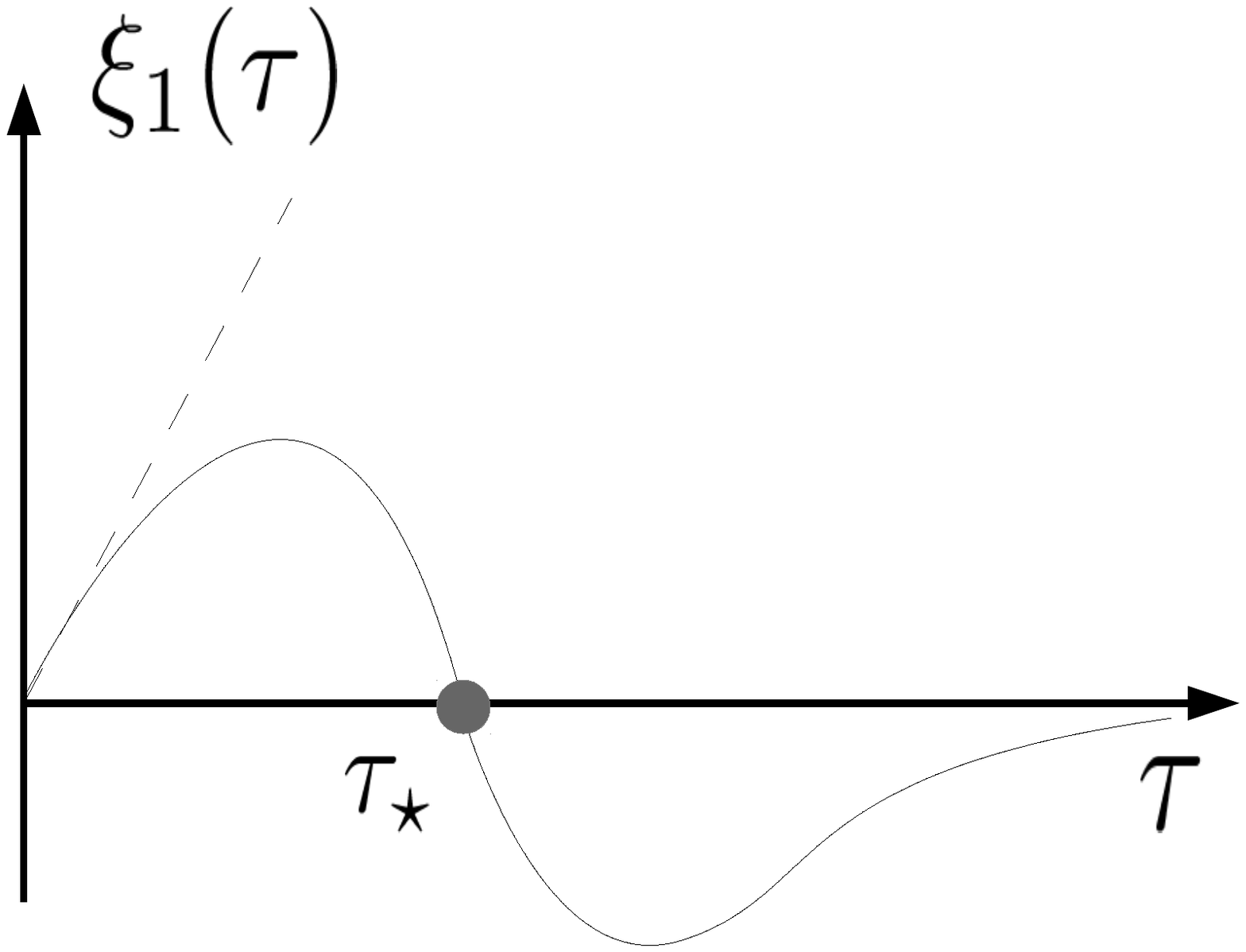} &
\includegraphics[trim = 0mm 120mm 0mm 15mm, scale=0.3]{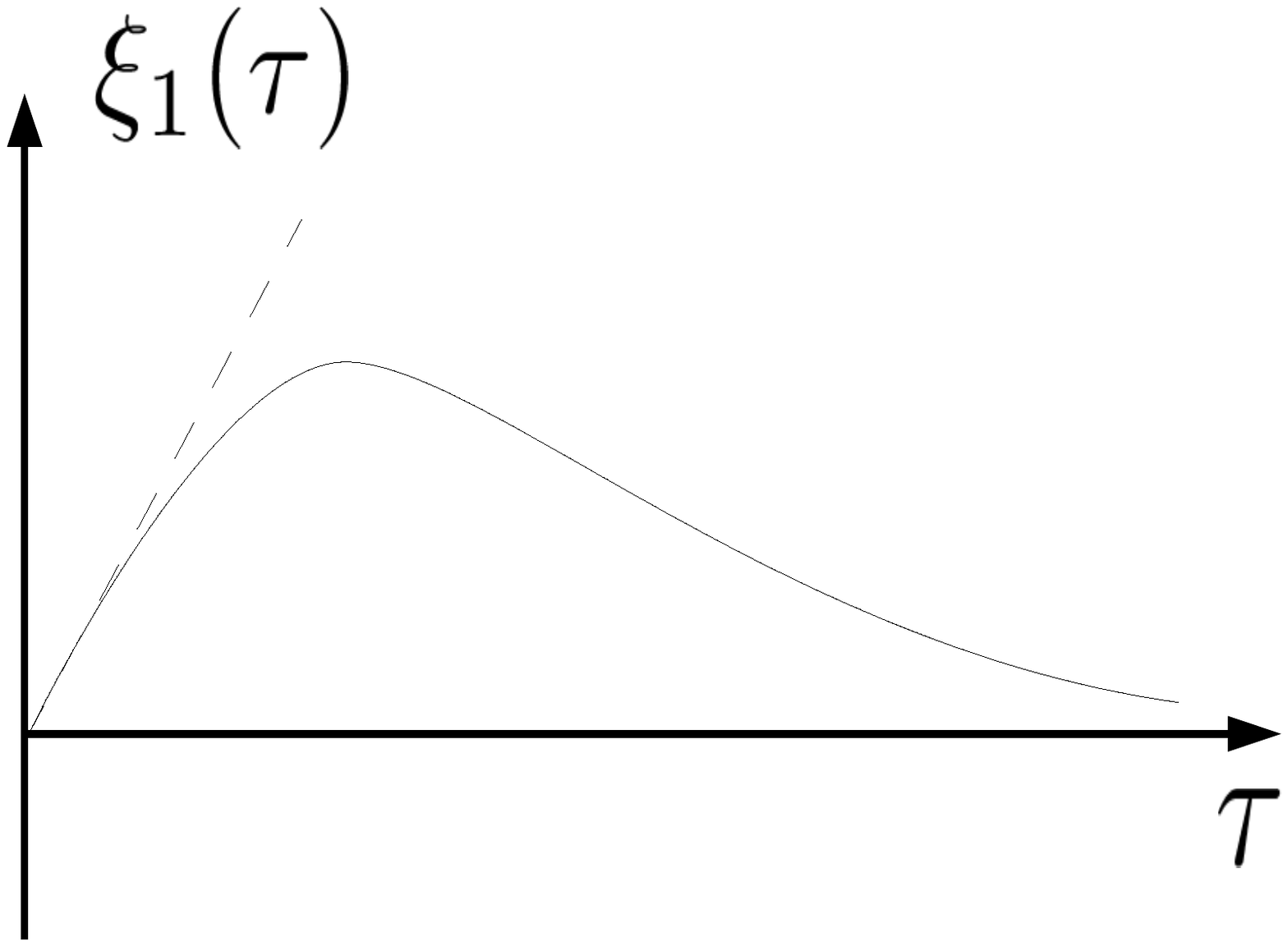}
\end{array}$
\caption{The function $\xi_1(\tau)$ is positive for small $\tau$ but cannot have a zero (\emph{left}) at finite $\tau=\tau_\star$, since $\dot{\xi}_1(\tau) < 0$ is not allowed. As a consequence, it will be everywhere positive (\emph{right}). Notice that it goes to zero at infinity, otherwise we do not get asymptotic KS solution.}
\label{xi1}
\end{center}
\end{figure}

As a consistency check we may show that the net force on a probe D3 brane in this background will be pointed towards the tip, as expected for a solution with smeared anti-D3 branes. This force is given by (\ref{forceD3}) and in our conventions it means that $\xi_1^+$ should be non-negative. Let us demonstrate it with the help of the $\xi_1^+$ equation of motion (we summarize the equations for $\xi_f^+$, $\xi_k^+$ and $\xi_F^+$ in Appendix \ref{DotXfkF}):
\begin{equation}
\label{dx1+eq}
    \dot{\xi}^+_1 - K e^{-2 x} \xi^+_1  =
      4 e^{2 x - 4(p + A)} \left[ e^{\Phi +2 y} (\xi_f^+)^2 + e^{\Phi- 2 y} (\xi_k^+)^2
        + \frac{1}{2} e^{-\Phi} (\xi_F^+)^2 \right] \, .
\end{equation}
We know from (\ref{IRexpansion}) that $\xi_1^+ = \frac{1}{2} c_0^2 \tau + \ldots$ near $\tau=0$. Thus, for function $\xi_1^+(\tau)$ to vanish at some $\tau=\tau_\star$ and to become negative for $\tau > \tau_\star$, we must have $\dot{\xi}^+_1(\tau_\star) < 0$. This, however, is at odds with the equation (\ref{dx1+eq}), since its right hand side is non-negative. We conclude that $\xi_1^+(\tau)>0$
for $\tau \in \left(0, \infty \right)$, see Figure \ref{xi1}.

Let us now come back to the $\xi^-$ equations of motion. We may further use (\ref{dx1-eq}) in order to extract a relation between $b_f$, $b_k$, $b_F$ and the constant $b_1$ defined by
\begin{equation}
    \xi^{-}_1 = b_1 \tau^2 + \mathcal{O} (\tau^3) \, .
\end{equation}
Plugging (\ref{bfkF}) into (\ref{dx1-eq}) we get:
\begin{equation}\label{b1bf}
b_1 = \frac{\pi Q}{3 c_0} \left( e^{\Phi_0} \left( \frac{b_f^2}{4}  + 4 b_k^2 \right) + e^{-\Phi_0}  \frac{b_F^2}{2} \right)
    = \frac{\pi Q}{12 c_0} e^{\Phi_0} \left( b_f^2  + 48 b_k^2 \right)
    \geqslant
    \frac{\pi Q}{12 c_0} e^{\Phi_0} b_f^2
\, .
\end{equation}
This last inequality will play a crucial r\^ole in the next section when we will determine the form of the polarization potential.
For this we will also need the explicit expressions for the RR 4 and 6-form gauge fields:\footnote{In our conventions $d C_6 = e^\Phi \star_{10} F_3 - H_3 \wedge C_4$.}
\begin{eqnarray} \label{C4C6}
  C_4 &=& \left( -2 \chi_1 + e^{-2 x + 4 (p+A)} \right) d x_0 \wedge \ldots \wedge d x_3 \, \nonumber\\
  C_6 &=& \chi_f \cdot g_1 \wedge g_2 + \chi_k \cdot g_3 \wedge g_4 \, ,
\end{eqnarray}
where\footnote{Notice that $C_6$ depends only on $\xi^{-}$'s and vanishes for the anti KS solution.  It is also zero for the KS background as one can show using (\ref{xi-eq}).}
\begin{equation}\label{chimodes}
    \dot{\chi}_1 = e^{-2 x} \xi^{-}_1 \, , \qquad
    \dot{\chi}_f = 4 e^{2 y + \Phi} \xi^{-}_f + 2 \dot{f} \chi_1 \, , \qquad
    \dot{\chi}_k = 4 e^{-2 y + \Phi} \xi^{-}_k + 2 \dot{k} \chi_1 \, .
\end{equation}
The integration constants of $\chi_1$, $\chi_f$ and $\chi_k$ can be eliminated by gauge transformations of $C_4$
and $C_6$. We will fix the freedom by requiring that all these functions vanish at $\tau=0$.
Using (\ref{bfkF}) and (\ref{IRexpansion}) we can find the leading order behavior of $\chi_1$ and $\chi_f$:
\begin{equation}
\label{IRchi}
\chi_1 = \frac{2}{\pi Q} b_1 \cdot \tau^2 + \ldots \ , \qquad
\chi_f = \frac{1}{3} e^{\Phi_0} b_f \cdot \tau^3 + \ldots \, .
\end{equation}

We end this section by making explicit the singular character of our solution, and explaining the various terms that contribute to the singularity. First, it is easy to verify by plugging the IR behavior into ~\eqref{SquaredH3F3} that the 3-form flux densities diverge, namely
\begin{equation}
\left\vert H_3 \right\vert^2 \sim  \frac{(b_f)^2 + 8 (b_k)^2}{\sqrt{\tau}} + {\cal O}(\tau^0) \, \quad \left\vert F_3 \right\vert^2\sim \frac{(b_k)^2}{\sqrt{\tau}} + {\cal O}(\tau^0) \ .
\end{equation}
It is useful to characterize the singularity of our solution in terms of the ISD and IASD components of the three form flux.
Borrowing the notation of \cite{Blaback:2012nf}, we define three scalar functions of the radial variable $\lambda^A$, by
\beq
e^{-\Phi} H_3 =  \lambda(\tau)_A * F_3^A
\eeq
where $F_3^A$ denotes each of the three components of $F_3$, namely along $g_{125}$, $g_{345}$ and $d\tau \, (g_{13}+g_{24})$. These definitions ensure that for  ISD (IASD) fluxes, $\lambda^A=-1 (1)$. We find that the component with legs $g_{345}$ is singular:
\beq
\lambda_{345}(\tau) =\frac{\pi Q}{2c_0 P} b_f \tau^{-1} + \mathcal{O}(\tau^0) \ ,
\eeq
while the other two are regular.
 We should note that in the linearized anti-D3 solution of~\cite{Bena:2009xk,Bena:2011hz,Bena:2011wh}, there was actually an additional singular $\lambda$, namely $\lambda_{125}$. We thus see that at the full non-linear level one singularity gets resolved, but the singularity in $\lambda_{345}$ persists, confirming the observation made in~\cite{Massai:2012jn}. Note that this corresponds precisely to three-form field strengths that have the legs on the $S^3$.

We end this section by comparing our results to the ones in the solution considered in~\cite{Blaback:2012nf}, corresponding to anti-D6-branes wrapping a $T^3$. As explained in~\cite{Bena:2012tx}, that solution can be T-dualized three times, and will yield a KS-like solution where the warped deformed conifold is replaced by $T^3 \times \mathbb{R}^3$. As argued in \cite{Bena:2012tx} this solution can be regarded as a toy model for the KS infrared region. Indeed, there is a flux singularity very much like the one found here, but in a sense simpler: the fluxes can be parameterized by a single function $\lambda$, defined by $H_3 = \lambda(\tau) *_3 F_0$, where $F_0$ is the mass parameter in massive type IIA, which is the toy-model version of the dual 3-form $F_3$ on the $S^3$. The fully backreacted anti-D6 solution has \cite{Blaback:2011pn} $\lambda(\tau) = \lambda_0 \, \tau^{-1} + \mathcal{O}(\tau^0)$, and the whole (IR singular) solution can be parameterized by $\lambda_0$.

\section{D5 polarization}
\label{D5polarization}

In this section we would like to address the main question of this paper: can the 3-form flux singularity of the anti-D3 brane putative solution be cured by the polarization of the anti-D3 branes into D5 branes? The singularity occurs at $\tau=0$ and if we find that there is a stable configuration with a polarized D5 brane wrapping the 2-sphere at a finite distance away from the tip, it will imply that the singularity is still physically meaningful. In the first subsection we will compute the potential of a probe D5 brane with anti-D3 charge $n$ in the singular solution sourced by $Q$ anti-D3 branes smeared on the KS tip. We will then argue in the second subsection that this potential also governs the polarization of {\em all} $Q$ anti-D3 branes into D5 branes.

\subsection{The D5 potential}

In order to see if the anti-branes polarize or not into D5-branes we need to compute the potential of a probe D5-brane that wraps the $S^2$ of the deformed conifold and has $n$ anti-D3 branes dissolved in it. The D5 brane action (in string frame) is
\begin{equation}
S_{D5} = S_{DBI} + S_{WZ} \,
\end{equation}
with
\begin{equation}
S_{DBI} = -\mu_5 \int \d^6 \xi
e^{-\phi} \sqrt{- \textrm{det} \left( g + 2 \pi \mathcal{F}_2 \right)}  \, ,
\quad
S_{WZ}  = \mu_5 \int \left( C_6 + 2 \pi \mathcal{F}_2 \wedge C_4 \right) \, ,
\end{equation}
where $2 \pi \mathcal{F}_2 \equiv 2 \pi \mathfrak{f}_2 - B_2$ and $\mathfrak{f}_2$ is the D5 worldvolume gauge field strength that gives the number of anti-D3 branes dissolved in the D5:
\begin{equation}
\mathfrak{f}_2 = \frac{n}{2} \omega_{S^2} \, ,
\end{equation}
where $\omega_{S^2}$ is proportional to $g_1 \wedge g_2$. The larger $n$ the easier to polarize it is, and in that limit one can expand the DBI action in a $1/n$ series. The leading term cancels the leading term in the WZ action, and the  polarization potential has in general the following form:
\begin{equation}
\label{Vtau}
       V(\tau) \sim  2 \pi n \cdot c_2 \tau^2 - c_3 \tau^3 + \frac{1}{2 \pi n} c_4 \tau^4 \, ,
\end{equation}
where the quadratic term comes from the imperfect WZ-DBI cancelation (and is equal to the force on a probe anti-D3 brane), the cubic term\footnote{In our conventions $c_3$ is positive.} comes from the $C_6$ term in the WZ action and the quartic terms is the subleading term in the $1/n$ expansion of the DBI action.

It is easy to show that if the following relation is satisfied
\begin{equation}\label{polbound}
    (c_3)^2 < \frac{32}{9}c_2 c_4  \, ,
\end{equation}
then the potential~\eqref{Vtau} has no minima for any $\tau$ away from zero, and thus there is no polarization.
In our singular solution we obtain
\begin{equation}
    c_2 = \lim_{\tau \to 0} \left( \frac{\chi_1}{\tau^2} \right) \, , \qquad
    c_3 = \lim_{\tau \to 0} \left( \frac{\chi_f}{\tau^3} \right) \, , \qquad
    c_4 = \lim_{\tau \to 0} \left( \frac{e^{4(p + A) + 2 y + \Phi}}{\tau^4} \right) \, ,
\end{equation}
where $\chi_1$ and $\chi_f$ are defined in eqs. (\ref{C4C6}), \eqref{chimodes}, and their IR behavior is given in \eqref{IRchi}. Using this and (\ref{IRexpansion}), we arrive at the following result:\footnote{Notice that neither $b_k$ nor $b_F$ appear in the potential.}
\begin{equation}
\label{c234}
c_2  = \frac{2}{\pi Q} b_1  \, , \qquad
c_3  = \frac{1}{3} e^{\Phi_0} b_f \, , \qquad
c_4  = \frac{1}{4} c_0  e^{\Phi_0} \, .
\end{equation}
We can now rewrite the inequality in (\ref{b1bf}) in terms of $c_2$, $c_3$ and $c_4$ and find that in all anti-D3 singular solutions:
\begin{equation}
\label{main}
    (c_3)^2 \leqslant \frac{8}{3} c_2 c_4  \, .
\end{equation}
From this result we see that the condition~\eqref{polbound} is always satisfied.
This is our main result. It proves that the potential (\ref{Vtau}) has no minimum, not even a metastable one. Thus \emph{no} polarization into D5 branes occurs and the 3-form flux singularity appears to be \emph{genuine}. Even more importantly, we were able to prove this statement without extending the solution from the IR all the way to the UV.

In fact, the story here is strikingly similar to the D6 toy model of \cite{Bena:2012tx} that we briefly mentioned in the previous section. Remarkably, in this model there is also no need to determine the full backreacted solution in order to see that the polarization potential has no minimum away from zero. Moreover, the inequality (\ref{main}) was exactly saturated. Our potential is more complicated, and reduces to the one of \cite{Bena:2012tx} if one sets $b_k=b_F=0$. However, turning this parameter back on makes polarization even \emph{more} difficult, and hence does not modify the physics that the toy model predicted.

\subsection{The mean field argument}
\label{TheMeanField}

To understand the relation between the potential for probe anti-D3 branes that we calculated in the previous section, the potential that governs the polarization of all the smeared D3 branes into smeared D5 branes, and the potential for the polarization of {\em localized} D3 branes into D5 branes it is important to recapitulate several very important features of the Polchinski-Strassler construction \cite{Polchinski:2000uf}.

Despite the absence of a fully-backreacted solution, Polchinski and Strassler  compute in \cite{Polchinski:2000uf} the potential for {\it all} the D3 branes that source the $AdS_5 \times S^5$ geometry to polarize into D5, NS5 or $(p,q)$-5 branes. This computation has three ingredients. One starts from a singular solution sourced by $N$ D3 branes, and calculates the potential of a probe D5 brane that wraps a topologically-trivial $S^2$ and has $n$ units of D3 brane charge inside, where $n \ll N$. This potential has three terms, that go respectively like $r^4, r^3$ and $r^2$. One then finds that in the $r^4$ term the various factors of the warp function of the backreacted D3 branes cancel out, and this term is therefore independent of the location of the backreacted branes (all the information about the angular location of the D branes is stored in the warp factor). Furthermore, the $r^3$ term is proportional to the IASD three-form, which is closed and co-closed, and hence depends only on the asymptotic boundary conditions; hence, this term is also independent of the location of the backreacted D3 branes.

The $r^2$ term in \cite{Polchinski:2000uf} is much more complicated, as it comes from the the backreaction of the fluxes on the metric, dilaton and five-form field strength. When supersymmetry is present, one can find this term by completing the squares in the supersymmetric polarization potential \cite{Polchinski:2000uf}. However, computing this term directly is much more painful, and has been done in \cite{Freedman:2000xb,Taylor:2001pp}. Not surprisingly, the two calculations agree, and the $r^2$ term also turns out to be independent of the warp factor sourced by the backreacted D3 branes, although this is much more difficult to see from the supergravity calculation. When supersymmetry is broken by the introduction of a fourth fermion mass, one can still compute the $r^2$ term by using various supersymmetric limits as well as the fact that this term comes from interacting three-form field strengths (see for example section IV of \cite{Polchinski:2000uf}) and one still finds that this term is independent of the warp factor, and therefore of the position of the backreacting D3 branes. Hence, both in supersymmetric and in non-supersymmetric Polchinski-Strassler backgrounds the polarization potential for a probe D5 brane with D3 charge $n$ is independent of the position of the $N$ D3 branes that source the solution.

Armed with this fact, one can consider then the much more general problem of a large number of D5 branes that have charges $n_i$, such that $\sum_i n_i=N$ and $n_i \ll N$. Each of these D5 branes can now be treated as a probe in the supergravity solution created by the other branes, and because the polarization potential is independent of the position of the D3 branes that source the background, the potential felt by each D5 brane in this configuration is the same as the potential of this D5 brane in the singular solution above. Hence, one can construct self-consistently the full solution by requiring  that each probe is at a minimum in the background sourced by the other probes. This ``mean-field'' construction can then be generalized straightforwardly to D3 branes polarizing into multiple shells that can also have NS5 or more general $(p,q)$-5-brane dipole charge. More generally, this construction can also be used to study all the other types of brane polarization that occur in the region where the branes that polarize dominate the geometry. The correctness of this ``mean-field" Polchinski-Strassler construction of vacua with polarized branes has been confirmed in the few examples where the fully-backreacted brane polarization supergravity solution exists, such as the mass-deformed M2 brane theory \cite{Bena:2004jw,Lin:2004nb}, or the supergravity dual of the mass-deformed 5D Super Yang-Mills theory \cite{Bena:2002wg}. Hence the probe calculation that we presented in the previous section gives the full potential for the smeared anti-D3 branes to polarize into D5 branes at a finite distance away from the tip.

However, one can do much more: one can use this independence of the Polchinski-Strassler polarization potential on the location of the polarizing branes to compute the potential for $N$ D3 branes that are {\it localized} near the north pole of the large $S^3$ at the bottom of the KS solution to polarize into a D5 brane wrapping the conifold $S^2$ at a finite distance from the tip. By the arguments above, this potential is the same as the potential for several probe D3 branes to polarize on this $S^2$ in the singular geometry sourced by a large number of D3 branes that are localized on the KS three-sphere, as long as the polarization occurs in the region where these D3 branes dominate the geometry. In turn, this potential is independent of the location of the D3 branes that dominate the geometry, and hence is the same as the potential for several probe D3's to polarize into a D5 brane in the geometry where these D3 branes are {\em smeared}, which we calculated in the previous subsection.

Hence, our calculation indicates that neither smeared nor localized anti-D3 branes do not polarize into D5 branes, and therefore that brane polarization \`a la Polchinski-Strassler does not appear to cure the singularity of antibranes in KS.

\subsection{Validity of approximations}

We now discuss the range of validity of our calculation. We see from
the probe D5 potential~\eqref{Vtau} and the expressions~\eqref{c234} that the radius $\tau_{\ast}$ at which the D5
would sit is of the order
\begin{equation}\label{taustar}
\tau_{\ast} \sim n \frac{c_3}{c_4} \sim n\, b_f \, .
 \end{equation}
Here we immediately face a problem. Since we do not know the full solution we cannot fix the dependence of $b_f$ and all other coefficients on $P$ (the 5-form flux) and $Q$ (the number of the anti D3's).  We can still \emph{estimate} however this dependence using the method utilized in \cite{Bena:2012tx}. The full solution is expected to be unique, namely having no parameters other than $P$ and $Q$. We, therefore, anticipate that for a fixed order in the $\tau$ expansion the contributions coming from various terms in the EOM will be of the same order in terms of $P$ and $Q$. In other words, there should be a detailed balance between different terms.

Let us introduce the following notation:
\begin{equation}
 \xi^-_f = b_f^{(0)} + b_f^{(1)} \tau + b_f^{(2)} \tau^2 + \ldots
\end{equation}
and similarly for the other $\xi^-$'s. The additional index stands for the power of $\tau$ and in terms of the notation introduced in the previous section we have $b_f^{(0)} = b_f$, $b_k^{(2)} = b_k$,  \emph{etc}.

We can start our analysis, for instance, from the $\tau^2$ contribution to the following term in the $\xi^-_1$ equation (\ref{dx1-eq})
\begin{equation}
    e^{2 y} \left( \xi^-_f \right)^2 + e^{-2 y} \left( \xi^-_k \right)^2 \,.
\end{equation}
We see that the detailed balance implies $b_f^{(0)} \sim b_k^{(2)}$. Next, the
$e^{2 y} \xi^-_f  - e^{-2 y} \xi^-_k $ term in the $\dot{\xi}^-_F$ equation gives $b_f^{(0)} \sim b_k^{(4)}$. We conclude that
$b_k^{(2)} \sim b_k^{(4)}$. With a bit of effort, one can further show that in fact all $b_f^{(i)}$'s and $b_k^{(i)}$'s are of the same order of magnitude. Moreover, $b_F^{(i)} \sim e^{\Phi_0} b_k^{(j)}$ for all $i$ and $j$.

Let us now consider the $b_1^{(i)}$ coefficients. From (\ref{dx1-eq}) and the $\dot{\xi}^-_f$ equation in (\ref{dxfkF-eq})
we learn that $b_1^{(2)} \sim Q e^{\Phi_0} \left( b_f^{(0)} \right)^2$ and $b_f^{(2)} \sim P Q^{-1} b_1^{(2)}$ respectively. Combining the two we see that $b_{f}^{(i)} \sim e^{-\Phi_0} P^{-1}$ and $b_{F}^{(i)} \sim P^{-1}$.

Finally, we have to compare the $\xi^{-}_f$ and ${\xi^{-}_f}^2$ terms on the right hand side of the $\dot{\xi^-_\Phi}$ equation in (\ref{DotTildeXi23yPhi}). We find $b_f^{(1)} \sim P/Q$ and comparing this with the observations of the previous two paragraphs we see eventually that $e^{\Phi_0} \sim Q/P^2 $.

To summarize, we find that:
\begin{equation}
\label{OrderOfMagnitude}
    b_{f}^{(i)} \sim b_{k}^{(i)} \sim \frac{P}{Q} \, , \qquad
    b_{F}^{(i)} \sim \frac{1}{P} \, , \qquad \textrm{and} \qquad
    e^{\Phi_0}  \sim \frac{Q}{P^2} \, .
\end{equation}
The remaining coefficients are irrelevant for our analysis.

We are now in a position to check the validity region of our polarization calculation.
In order to trust our computation we need to assume the following conditions:
\begin{itemize}
    \item The anti-D3 charge of the probe D5 should be much smaller than the anti-D3 charge of the background
    \begin{equation}\label{Qbiggern}
        n  \ll Q \, .
    \end{equation}
    We recall that in our conventions $Q$ is the number of anti-D3 branes.
    \item In order to trust the IR expansions we should demand that
     $\tau_{\ast}$ is small compared to the ratio between the leading and next-to-leading terms in the series. Since, for instance, all of the $b_f^{(i)}$ are of the same order, we must require $\tau_{\ast} \ll 1$. This in turn amounts to
    \begin{equation}
      n \ll \frac{Q}{P} \, .
    \end{equation}
    \item Since we expanded the square root in the DBI action we should
      demand that $\det(2\pi \mathcal{F}_2) \gg \det (g_{\perp})$. Recalling that in our Ansatz
         $\det (g_{\perp}) = e^{2x+2y} \sim Q \tau^3 + \ldots$ we obtain $n^2 \gg \tau_{\ast}^3 Q$ or
    \begin{equation}\label{DBIexpand}
        n \ll \frac{Q^2}{P^3}  \, .
    \end{equation}
    \item The radius of the $S^2$ at which the D5 brane would polarize should be large in string units. Since the radius is given by $(\det (g_{\perp}))^{1/4}$ this amounts to demanding $\tau_{\ast}^{3} Q \gg 1$ or
    \begin{equation}\label{largeshpere}
        n \gg \frac{Q^{2/3}}{P} \, .
    \end{equation}
    \item The string coupling should be small at $\tau_{\ast}$. This means that
        \begin{equation}
            Q \ll P^2 \, .
        \end{equation}
\end{itemize}
To conclude, we have the following criteria
\begin{equation}
\label{thelimit}
    \frac{Q^{2/3}}{P} \ll n \ll \frac{Q^2}{P^3} \ll \frac{Q}{P} \ll Q \, .
\end{equation}
This can be easily achieved. For example, we can set $n \sim \sigma$, $Q \sim \sigma^7$  and $P \sim \sigma^4$ for large $\sigma$.

Before closing this section let us add an important comment on the range of the parameters. As one can see from (\ref{thelimit}) our calculation necessitates $Q \gg P$, in other words the number of the anti D3's has to dominate the flux. At the first glance it looks as we are away from the parameter region studied in \cite{Kachru:2002gs}, and so our findings have nothing to do with the brane polarization scenario proposed in this paper.

Let us clarify this important point. Indeed, the probe analysis of \cite{Kachru:2002gs} indicates that if the ratio $Q/P$ is below the threshold of approximately $8\%$, then the polarization potential has a metastable vacuum, so that the 2-sphere warped by the polarizing NS5 brane can be stabilized at a certain radius inside the 3-sphere at the conifold tip. If the ratio rather exceeds the threshold, the polarization potential is monotonic and there is only one (supersymmetric) vacuum. In other words, the probe anti brane polarizes for any values of $Q$ and $P$, but only for small enough $Q/P$ the configuration has a non-supersymmetric metastable vacuum, so that the anti branes do not dissolve directly into the flux. One might expect that once the backreaction is taken into account the threshold will be lower, but nevertheless the scenario will still work for small enough $Q/P$.

Furthermore, one may also argue \cite{Bena:2012tx} that if one considers the polarization of the anti D3 branes into multiple NS5 branes, one might see that there will exist metastable vacua for an arbitrarily-large $Q$, as long as $Q/P$ divided by the number of these NS5 branes does not exceed $8\%$.

What we observe in this paper is conceptually different. We consider a different 2-sphere, one that shrinks at the conifold tip, and find that the anti D3's do not polarize into D5's on this 2-spere, since the polarization potential has neither metastable nor any other minimum at a finite distance away from the tip. Moreover, as no polarization occurs for large $Q/P$, it becomes even more unlikely for small $Q/P$. This is radically different from the situation in \cite{Kachru:2002gs}.

To sum up, the restriction we find on the ratio $Q/P$ does not invalidate in any way our main conclusions.

\section{Discussion}

We reviewed in detail the solution corresponding to $Q$ anti-D3-branes smeared on the $S^3$ at the tip of the deformed conifold, focusing on the fact that such solution has singular three-form fluxes in the infrared. These singularities could have suggested a stringy resolution by polarization \`a la Polchinski-Strassler. However, we show in this paper that the anti-D3-branes do not polarize into anti-D5-branes wrapping the $S^2$ at a finite radius, and therefore such mechanism of resolution of singularities is not in place here.

In order to show that, we computed the polarization potential, which has quadratic, cubic and quartic terms in the radial variable, but with coefficients such that there is no minimum, regardless of any UV data. All information needed to reach that conclusion are the IR boundary conditions reviewed in detail in the text. This result is quite strong, as on one hand we had shown that any solution with anti-D3-brane boundary conditions leads to either an anti-KS solution or to a singular solution, and on the other hand we are showing that this singularity is not resolved by polarization into anti-D5-branes, no matter what irrelevant or relevant operators one adds in the UV.

It is worth mentioning again the striking similarities between our results and those on anti-D6-branes in backgrounds with D6-charge dissolved in fluxes, which serves indeed as a toy model for the IR of KS. They both have the same type of singularities, and in neither case these can be resolved by polarizing into anti-branes of two dimensions higher. Furthermore, the potential for polarization in the case of anti-D3 branes reduces exactly to the one for anti-D6 if one integration constant is set to zero. The second integration constant, which should be related to the first one by UV boundary conditions, only makes things worse in terms of getting a minimum.

Our result also suggests that in the fully back-reacted solution there will be no polarization into NS5-branes, opposite to what happens in the probe calculation. In order to pin down this question one would need the localized solution, though, as smearing the charge on the $S^3$ wipes out this polarization channel. However, one might hope that, as was the case here, only very few details of the solution are needed to get an answer, and such details might be within reach. We hope to report on this soon.

\subsection*{Acknowledgements} We would like to thank Matteo Bertolini, Anatoly Dymarsky, Liam McAllister, Paul McGuirk and Timm Wrase for discussions, and specially Thomas Van Riet for stimulating exchanges and Dutch grammar corrections. This work was supported in part by the ANR grant 08-JCJC-0001-0 and the ERC Starting Grants 240210 - String-QCD-BH and 259133 -- ObservableString. SM is supported in part by a Contrat de Formation par la Recherche of CEA/Saclay. SK would like to express his deep discontent over the EURO 2012 schedule that caused numerous disturbances and delays over the course of this work.

\appendix

\section{The KS solution}
\label{AppendixKS}

Here we summarize all the functions of the KS solution as it appears in (\ref{setup}). The flux and the dilaton functions are given by
\begin{eqnarray}
\label{KS1}
    f(\tau) &=& - g_s P \cdot(\tau \coth (\tau) - 1) \tanh \left(\frac{\tau}{2} \right) \nonumber \\
    k(\tau) &=& - g_s P \cdot(\tau \coth (\tau) - 1) \coth \left(\frac{\tau}{2} \right) \nonumber \\
    F(\tau) &=& P \cdot \left( 1 - \frac{\tau}{\sinh (\tau)} \right) \nonumber \\
    e^{\Phi} &=& g_s \, .
\end{eqnarray}
The metric functions are
\begin{eqnarray}
\label{KS2}
   e^{2 x(\tau)} &=& \frac{h(\tau)}{16} \left( \frac{1}{2} \sinh(2 \tau) - \tau \right)^{2/3}  \nonumber \\
   e^{y(\tau)}   &=& \tanh \left( \frac{\tau}{2} \right)  \nonumber \\
   e^{6 p(\tau)} &=& \frac{24}{h(\tau)} \frac{\left( \frac{1}{2} \sinh(2 \tau) - \tau \right)^{1/3}}{\sinh^2(\tau)}  \nonumber \\
   e^{6 A(\tau)} &=&
        \frac{\epsilon_0^4}{3 \cdot 2^9} h(\tau) \left( \frac{1}{2} \sinh(2 \tau) - \tau \right)^{2/3} \sinh^2(\tau)  \, .
\end{eqnarray}

\section{The $\xi^{-}_2$, $\xi^{-}_3$, $\xi^{-}_y$ and $\xi^{-}_\Phi$ EOMs }
\label{DotTildeXi}

The equations of motion for the eight $\xi^{-}_a$ modes are given in equations~\eqref{dx1-eq}-\eqref{dxfkF-eq}, together with the following remaining equations:
\begin{eqnarray}
\label{DotTildeXi23yPhi}
  \dot{\xi}^{-}_2    &=&  - K e^{-2 x} \xi^{-}_1 + 3 e^{-6 p - 2 x}  \xi^{-}_2  \nonumber\\
                      && \qquad   - e^{- 4(p + A)} \left( (\xi^{-}_1)^2 + \frac{2}{3} \xi^{-}_2 \xi^{-}_3 -
                     \frac{1}{18} (\xi^{-}_3)^2 + 2 (\xi^{-}_y)^2 + 4 (\xi^{-}_\Phi)^2 \right) \nonumber\\
  \dot{\xi}^{-}_3    &=& 6 e^{- 6p - 2x} \xi^{-}_2 \nonumber\\
  \dot{\xi}^{-}_y    &=& \cosh y \cdot \xi^{-}_y + \frac{1}{3} \sinh y \cdot \xi^{-}_3
                           - 2 e^\Phi \left( e^{2 y} (2 P - F) \xi^{-}_f - e^{-2 y} F \xi^{-}_k \right) \nonumber\\
                      &&  + 4 e^{\Phi + 2 x - 4 (p + A)} \left( e^{2 y} (\xi^{-}_f)^2 - e^{-2 y} (\xi^{-}_k)^2 \right) \nonumber\\
  \dot{\xi}^{-}_\Phi &=& - e^\Phi \left( e^{2 y} (2 P - F) \xi^{-}_f + e^{-2 y} F \xi^{-}_k \right) +
                             \frac{1}{2} e^{-\Phi} (k - f) \xi^{-}_F                                       \nonumber\\
                        && + 2 e^{2 x - 4 (p + A)} \left( e^{\Phi} \left( e^{2 y} (\xi^{-}_f)^2 + e^{-2 y} (\xi^{-}_k)^2 \right)
                                -\frac{1}{2} e^{-\Phi} (\xi^{-}_F)^2 \right)  \, .
\end{eqnarray}

\section{The $\dot{\xi}^+_a$ equations}
\label{DotXfkF}

\begin{eqnarray}
    \dot{\xi}^+_f &=& - \frac{1}{2} e^{-2 x} (2 P - F) \xi_1^+ - \frac{1}{2} e^{-\Phi} \xi_F^+    \nonumber \\
    \dot{\xi}^+_k &=& - \frac{1}{2} e^{-2 x} F \xi_1^+ + \frac{1}{2} e^{-\Phi} \xi_F^+   \\
    \dot{\xi}^+_F &=& - \frac{1}{2} e^{-2 x} (k - f) \xi_1^+ -
                    e^{\Phi} \left( e^{2 y} \xi_f^+ - e^{-2 y} \xi_k^+ \right)   \, .   \nonumber
\end{eqnarray}

\bibliographystyle{utphys}
\bibliography{PSforKS}

\providecommand{\href}[2]{#2}\begingroup\raggedright\begin{thebibliography}{10}

\bibitem{Klebanov:2000hb}
I.~R. Klebanov and M.~J. Strassler, ``{Supergravity and a confining gauge
  theory: Duality cascades and $\chi$SB resolution of naked singularities},''
  {\em JHEP} {\bf 0008} (2000) 052,
  \href{http://xxx.lanl.gov/abs/hep-th/0007191}{{\tt hep-th/0007191}}.

\bibitem{Kachru:2002gs}
S.~Kachru, J.~Pearson, and H.~L. Verlinde, ``{Brane / flux annihilation and the
  string dual of a nonsupersymmetric field theory},'' {\em JHEP} {\bf 0206}
  (2002) 021, \href{http://xxx.lanl.gov/abs/hep-th/0112197}{{\tt
  hep-th/0112197}}.

\bibitem{Bena:2009xk}
I.~Bena, M.~Grana, and N.~Halmagyi, ``{On the Existence of Meta-stable Vacua in
  Klebanov-Strassler},'' {\em JHEP} {\bf 1009} (2010) 087,
  \href{http://xxx.lanl.gov/abs/0912.3519}{{\tt 0912.3519}}.

\bibitem{Bena:2011hz}
I.~Bena, G.~Giecold, M.~Grana, N.~Halmagyi, and S.~Massai, ``{On Metastable
  Vacua and the Warped Deformed Conifold: Analytic Results},'' {\em
  Class.Quant.Grav.} {\bf 30} (2013) 015003,
  \href{http://xxx.lanl.gov/abs/1102.2403}{{\tt 1102.2403}}.

\bibitem{Bena:2011wh}
I.~Bena, G.~Giecold, M.~Grana, N.~Halmagyi, and S.~Massai, ``{The backreaction
  of anti-D3 branes on the Klebanov-Strassler geometry},''
  \href{http://xxx.lanl.gov/abs/1106.6165}{{\tt 1106.6165}}.

\bibitem{Dymarsky:2011pm}
A.~Dymarsky, ``{On gravity dual of a metastable vacuum in Klebanov-Strassler
  theory},'' {\em JHEP} {\bf 1105} (2011) 053,
  \href{http://xxx.lanl.gov/abs/1102.1734}{{\tt 1102.1734}}.

\bibitem{Bena:2012bk}
I.~Bena, M.~Grana, S.~Kuperstein, and S.~Massai, ``{Anti-D3's - Singular to the
  Bitter End},'' \href{http://xxx.lanl.gov/abs/1206.6369}{{\tt 1206.6369}}.

\bibitem{Girardello:1999bd}
L.~Girardello, M.~Petrini, M.~Porrati, and A.~Zaffaroni, ``{The Supergravity
  dual of N=1 superYang-Mills theory},'' {\em Nucl.Phys.} {\bf B569} (2000)
  451--469, \href{http://xxx.lanl.gov/abs/hep-th/9909047}{{\tt
  hep-th/9909047}}.

\bibitem{Polchinski:2000uf}
J.~Polchinski and M.~J. Strassler, ``{The String dual of a confining
  four-dimensional gauge theory},''
  \href{http://xxx.lanl.gov/abs/hep-th/0003136}{{\tt hep-th/0003136}}.

\bibitem{McGreevy:2000cw}
J.~McGreevy, L.~Susskind, and N.~Toumbas, ``{Invasion of the giant gravitons
  from Anti-de Sitter space},'' {\em JHEP} {\bf 0006} (2000) 008,
  \href{http://xxx.lanl.gov/abs/hep-th/0003075}{{\tt hep-th/0003075}}.

\bibitem{Lin:2004nb}
H.~Lin, O.~Lunin, and J.~M. Maldacena, ``{Bubbling AdS space and 1/2 BPS
  geometries},'' {\em JHEP} {\bf 0410} (2004) 025,
  \href{http://xxx.lanl.gov/abs/hep-th/0409174}{{\tt hep-th/0409174}}.

\bibitem{Klebanov:2000nc}
I.~R. Klebanov and A.~A. Tseytlin, ``{Gravity duals of supersymmetric SU(N) x
  SU(N+M) gauge theories},'' {\em Nucl.Phys.} {\bf B578} (2000) 123--138,
  \href{http://xxx.lanl.gov/abs/hep-th/0002159}{{\tt hep-th/0002159}}.

\bibitem{Bena:2012ek}
I.~Bena, A.~Buchel, and O.~J. Dias, ``{Horizons cannot save the Landscape},''
  {\em Phys.Rev.} {\bf D87} (2013) 063012,
  \href{http://xxx.lanl.gov/abs/1212.5162}{{\tt 1212.5162}}.

\bibitem{Myers:1999ps}
R.~C. Myers, ``{Dielectric branes},'' {\em JHEP} {\bf 9912} (1999) 022,
  \href{http://xxx.lanl.gov/abs/hep-th/9910053}{{\tt hep-th/9910053}}.

\bibitem{Krishnan:2008gx}
C.~Krishnan and S.~Kuperstein, ``{The Mesonic Branch of the Deformed
  Conifold},'' {\em JHEP} {\bf 0805} (2008) 072,
  \href{http://xxx.lanl.gov/abs/0802.3674}{{\tt 0802.3674}}.

\bibitem{Pufu:2010ie}
S.~S. Pufu, I.~R. Klebanov, T.~Klose, and J.~Lin, ``{Green's Functions and
  Non-Singlet Glueballs on Deformed Conifolds},'' {\em J.Phys.} {\bf A44}
  (2011) 055404, \href{http://xxx.lanl.gov/abs/1009.2763}{{\tt 1009.2763}}.

\bibitem{Dymarsky:2005xt}
A.~Dymarsky, I.~R. Klebanov, and N.~Seiberg, ``{On the moduli space of the
  cascading SU(M+p) x SU(p) gauge theory},'' {\em JHEP} {\bf 0601} (2006) 155,
  \href{http://xxx.lanl.gov/abs/hep-th/0511254}{{\tt hep-th/0511254}}.

\bibitem{Kachru:2003aw}
S.~Kachru, R.~Kallosh, A.~D. Linde, and S.~P. Trivedi, ``{De Sitter vacua in
  string theory},'' {\em Phys.Rev.} {\bf D68} (2003) 046005,
  \href{http://xxx.lanl.gov/abs/hep-th/0301240}{{\tt hep-th/0301240}}.

\bibitem{Saltman:2004sn}
A.~Saltman and E.~Silverstein, ``{The Scaling of the no scale potential and de
  Sitter model building},'' {\em JHEP} {\bf 0411} (2004) 066,
  \href{http://xxx.lanl.gov/abs/hep-th/0402135}{{\tt hep-th/0402135}}.

\bibitem{Lebedev:2006qq}
O.~Lebedev, H.~P. Nilles, and M.~Ratz, ``{De Sitter vacua from matter
  superpotentials},'' {\em Phys.Lett.} {\bf B636} (2006) 126--131,
  \href{http://xxx.lanl.gov/abs/hep-th/0603047}{{\tt hep-th/0603047}}.

\bibitem{Balasubramanian:2005zx}
V.~Balasubramanian, P.~Berglund, J.~P. Conlon, and F.~Quevedo, ``{Systematics
  of moduli stabilisation in Calabi-Yau flux compactifications},'' {\em JHEP}
  {\bf 0503} (2005) 007, \href{http://xxx.lanl.gov/abs/hep-th/0502058}{{\tt
  hep-th/0502058}}.

\bibitem{Rummel:2011cd}
M.~Rummel and A.~Westphal, ``{A sufficient condition for de Sitter vacua in
  type IIB string theory},'' {\em JHEP} {\bf 1201} (2012) 020,
  \href{http://xxx.lanl.gov/abs/1107.2115}{{\tt 1107.2115}}.

\bibitem{Bena:2012tx}
I.~Bena, D.~Junghans, S.~Kuperstein, T.~Van~Riet, T.~Wrase, {\em et.~al.},
  ``{Persistent anti-brane singularities},'' {\em JHEP} {\bf 1210} (2012) 078,
  \href{http://xxx.lanl.gov/abs/1205.1798}{{\tt 1205.1798}}.

\bibitem{Papadopoulos:2000gj}
G.~Papadopoulos and A.~A. Tseytlin, ``{Complex geometry of conifolds and
  five-brane wrapped on two sphere},'' {\em Class.Quant.Grav.} {\bf 18} (2001)
  1333--1354, \href{http://xxx.lanl.gov/abs/hep-th/0012034}{{\tt
  hep-th/0012034}}.

\bibitem{Butti:2004pk}
A.~Butti, M.~Grana, R.~Minasian, M.~Petrini, and A.~Zaffaroni, ``{The Baryonic
  branch of Klebanov-Strassler solution: A supersymmetric family of SU(3)
  structure backgrounds},'' {\em JHEP} {\bf 0503} (2005) 069,
  \href{http://xxx.lanl.gov/abs/hep-th/0412187}{{\tt hep-th/0412187}}.

\bibitem{Candelas:1989js}
P.~Candelas and X.~C. de~la Ossa, ``{Comments on Conifolds},'' {\em Nucl.Phys.}
  {\bf B342} (1990) 246--268.

\bibitem{Kuperstein:2003yt}
S.~Kuperstein and J.~Sonnenschein, ``{Analytic nonsupersymmtric background dual
  of a confining gauge theory and the corresponding plane wave theory of
  hadrons},'' {\em JHEP} {\bf 0402} (2004) 015,
  \href{http://xxx.lanl.gov/abs/hep-th/0309011}{{\tt hep-th/0309011}}.

\bibitem{Grana:2000jj}
M.~Grana and J.~Polchinski, ``{Supersymmetric three form flux perturbations on
  AdS(5)},'' {\em Phys.Rev.} {\bf D63} (2001) 026001,
  \href{http://xxx.lanl.gov/abs/hep-th/0009211}{{\tt hep-th/0009211}}.

\bibitem{Gubser:2000vg}
S.~S. Gubser, ``{Supersymmetry and F theory realization of the deformed
  conifold with three form flux},''
  \href{http://xxx.lanl.gov/abs/hep-th/0010010}{{\tt hep-th/0010010}}.

\bibitem{Borokhov:2002fm}
V.~Borokhov and S.~S. Gubser, ``{Nonsupersymmetric deformations of the dual of
  a confining gauge theory},'' {\em JHEP} {\bf 0305} (2003) 034,
  \href{http://xxx.lanl.gov/abs/hep-th/0206098}{{\tt hep-th/0206098}}.

\bibitem{Bena:2010gs}
I.~Bena, G.~Giecold, and N.~Halmagyi, ``{The Backreaction of Anti-M2 Branes on
  a Warped Stenzel Space},'' {\em JHEP} {\bf 1104} (2011) 120,
  \href{http://xxx.lanl.gov/abs/1011.2195}{{\tt 1011.2195}}.

\bibitem{Massai:2011vi}
S.~Massai, ``{Metastable Vacua and the Backreacted Stenzel Geometry},'' {\em
  JHEP} {\bf 1206} (2012) 059, \href{http://xxx.lanl.gov/abs/1110.2513}{{\tt
  1110.2513}}.

\bibitem{Giecold:2011gw}
G.~Giecold, E.~Goi, and F.~Orsi, ``{Assessing a candidate IIA dual to
  metastable supersymmetry-breaking},'' {\em JHEP} {\bf 1202} (2012) 019,
  \href{http://xxx.lanl.gov/abs/1108.1789}{{\tt 1108.1789}}.

\bibitem{Blaback:2010sj}
J.~Blaback, U.~H. Danielsson, D.~Junghans, T.~Van~Riet, T.~Wrase, {\em
  et.~al.}, ``{Smeared versus localised sources in flux compactifications},''
  {\em JHEP} {\bf 1012} (2010) 043,
  \href{http://xxx.lanl.gov/abs/1009.1877}{{\tt 1009.1877}}.

\bibitem{Blaback:2011nz}
J.~Blaback, U.~H. Danielsson, D.~Junghans, T.~Van~Riet, T.~Wrase, {\em
  et.~al.}, ``{The problematic backreaction of SUSY-breaking branes},'' {\em
  JHEP} {\bf 1108} (2011) 105, \href{http://xxx.lanl.gov/abs/1105.4879}{{\tt
  1105.4879}}.

\bibitem{Blaback:2011pn}
J.~Blaback, U.~H. Danielsson, D.~Junghans, T.~Van~Riet, T.~Wrase, {\em
  et.~al.}, ``{(Anti-)Brane backreaction beyond perturbation theory},'' {\em
  JHEP} {\bf 1202} (2012) 025, \href{http://xxx.lanl.gov/abs/1111.2605}{{\tt
  1111.2605}}.

\bibitem{Blaback:2012nf}
J.~Blaback, U.~H. Danielsson, and T.~Van~Riet, ``{Resolving anti-brane
  singularities through time-dependence},''
  \href{http://xxx.lanl.gov/abs/1202.1132}{{\tt 1202.1132}}.

\bibitem{Massai:2012jn}
S.~Massai, ``{A Comment on anti-brane singularities in warped throats},''
  \href{http://xxx.lanl.gov/abs/1202.3789}{{\tt 1202.3789}}.

\bibitem{Freedman:2000xb}
D.~Z. Freedman and J.~A. Minahan, ``{Finite temperature effects in the
  supergravity dual of the N=1* gauge theory},'' {\em JHEP} {\bf 0101} (2001)
  036, \href{http://xxx.lanl.gov/abs/hep-th/0007250}{{\tt hep-th/0007250}}.

\bibitem{Taylor:2001pp}
M.~Taylor, ``{Anomalies, counterterms and the N=0 Polchinski-Strassler
  solutions},'' \href{http://xxx.lanl.gov/abs/hep-th/0103162}{{\tt
  hep-th/0103162}}.

\bibitem{Bena:2004jw}
I.~Bena and N.~P. Warner, ``{A Harmonic family of dielectric flow solutions
  with maximal supersymmetry},'' {\em JHEP} {\bf 0412} (2004) 021,
  \href{http://xxx.lanl.gov/abs/hep-th/0406145}{{\tt hep-th/0406145}}.

\bibitem{Bena:2002wg}
I.~Bena and C.~Ciocarlie, ``{Exact N=2 supergravity solutions with polarized
  branes},'' {\em Phys.Rev.} {\bf D70} (2004) 086005,
  \href{http://xxx.lanl.gov/abs/hep-th/0212252}{{\tt hep-th/0212252}}.

\end{thebibliography}\endgroup

\end{document}